\documentclass[acmsmall]{acmart}
\usepackage{array}
\usepackage{amsmath}
\usepackage{multirow}
\usepackage{multicol}
\usepackage{enumitem}
\usepackage{caption}
\usepackage{makecell}
\usepackage{colortbl}
\usepackage{pifont}
\usepackage{xcolor}
\usepackage{tcolorbox}
\usepackage{listings}
\usepackage{hyperref} 
\usepackage{footnote}
\newcommand{\cmark}{\textcolor{black}{\ding{51}}} 
\newcommand{\xmark}{\textcolor{gray}{\ding{55}}} 

\usepackage{cellspace}
\AtBeginDocument{%
  }

\definecolor{codebg}{rgb}{0.95,0.95,0.95}
\definecolor{keywordcolor}{rgb}{0.15,0.32,0.87}
\definecolor{commentcolor}{rgb}{0.25,0.6,0.35}
\definecolor{stringcolor}{rgb}{0.76,0.14,0.18}
\definecolor{funccolor}{rgb}{0.76,0.24,0.10}
\definecolor{darkgreen}{rgb}{0.0,0.5,0.0}
\definecolor{darkyellow}{rgb}{0.6,0.5,0.0}
\definecolor{lightred}{rgb}{1.0,0.5,0.5}
\lstdefinestyle{mystyle}{
 backgroundcolor=\color{codebg},
    basicstyle=\ttfamily\tiny,
    breaklines=true,
    keywordstyle=\color{keywordcolor},
    stringstyle=\color{stringcolor},
    commentstyle=\color{commentcolor},
    tabsize=4,
    showspaces=false,
    showstringspaces=false,
    language=C++,
    morekeywords={alignas, alignof, and, and_eq, asm, auto, bitand, bitor, bool, break, case, catch, char, char16_t, char32_t, class, compl, const, constexpr, const_cast, continue, decltype, default, delete, do, double, dynamic_cast, else, enum, explicit, export, extern, false, float, for, friend, goto, if, inline, int, long, mutable, namespace, new, noexcept, not, not_eq, nullptr, operator, or, or_eq, private, protected, public, register, reinterpret_cast, return, short, signed, sizeof, static, static_assert, static_cast, struct, switch, template, this, thread_local, throw, true, try, typedef, typeid, typename, union, unsigned, using, virtual, void, volatile, wchar_t, while, xor, xor_eq},
    emph={thread1, thread2, main, assert,pthread, atomic},
    emphstyle=\color{funccolor},
}

\renewcommand{\arraystretch}{-100.5}

\begin{document}

\title{Assessing Large Language Models in Comprehending and Verifying Concurrent Programs across Memory Models}

\author{Ridhi Jain}
\affiliation{%
  \institution{IIIT-Delhi}
  \country{India}}
\email{ridhij@iiitd.ac.in}

\author{Rahul Purandare}
\affiliation{%
  \institution{University of Nebraska--Lincoln}
  \country{USA}}
\email{rahul@unl.edu}

\renewcommand{\shortauthors}{R. Jain and R. Purandare}

\begin{abstract}
As concurrent programming becomes increasingly prevalent, effectively identifying and addressing concurrency issues such as data races and deadlocks is critical. This study evaluates the performance of several leading large language models (LLMs), including GPT-3.5-turbo, GPT-4, GPT-4o, GPT-4o-mini, and Mistral-AI's Large2, in understanding and analyzing concurrency issues within software programs. Given that relaxed memory models, such as Total Store Order (TSO) and Partial Store Order (PSO), are widely implemented and adapted in modern systems—supported even by commodity architectures like ARM and x86—our evaluation focuses not only on sequentially consistent memory models but also on these relaxed memory models. Specifically, we assess two main aspects: the models' capacity to detect concurrency problems under a sequentially consistent memory model and their ability to verify the correctness conditions of concurrent programs across both sequentially consistent and relaxed memory models. To do this, we leverage SV-COMP's \textit{pthread} tests and 25 ARM Litmus tests designed to evaluate Total Store Order (TSO) and Partial Store Order (PSO) memory models.

The experimental results reveal that GPT-4, GPT-4o, and Mistral-AI's Large2 demonstrate a robust understanding of concurrency issues, effectively identifying data races and deadlocks when assessed under a sequentially consistent memory model. However, despite its superior performance, all selected LLMs face significant challenges verifying program correctness under relaxed memory models. These LLMs exhibit limitations in accurately capturing memory ordering constraints, and their current capabilities fall short in verifying even small programs in these complex scenarios. This study highlights the gap in LLMs' effectiveness for relaxed memory models and emphasizes the need for further advancements to improve their reliability in concurrent program analysis and verification.
\end{abstract}



\maketitle

\section{Introduction}
\label{introduction}

As modern software systems grow increasingly complex and hardware evolves to feature multi-core processors, the ability to run multiple threads or processes concurrently has become essential. This shift toward concurrency is driven by the need to maximize performance and optimize resource utilization. However, along with these benefits comes significant challenges: ensuring that concurrent programs operate correctly. Issues like race conditions, deadlocks, and synchronization errors can lead to unpredictable behavior and critical failures in software systems. These complexities make verifying and testing concurrent programs one of the most difficult tasks in software development. These issues are further compounded by the intricacies introduced by relaxed memory models, which define how operations on shared memory (reads and writes) are ordered and observed by different threads or processors in a concurrent system.

Traditional software engineering techniques like static and dynamic analysis have long been employed to detect bugs, optimize performance, and verify program correctness. Static analysis examines code without executing it, identifying issues such as memory leaks or null pointer dereferences by analyzing control flow, data flow, and structure. These techniques rely on predefined rules and heuristics, often tailored to specific programming environments. In contrast, dynamic analysis monitors a program during execution to catch runtime issues like race conditions and deadlocks that static analysis might miss. While both methods are effective, they typically demand extensive domain knowledge and significant computational resources, particularly when applied to large or complex codebases. Running static analysis on massive projects or conducting dynamic analysis on intricate systems can be computationally expensive, often resulting in long processing times. Moreover, their reliance on predefined rules makes them less adaptable to rapidly evolving software environments, limiting their flexibility in modern, agile development contexts.

Deep learning techniques are increasingly adopted in software engineering and testing to address these challenges. By leveraging models such as recurrent neural networks (RNNs)~\cite{fan2019software, liu2022deepstate}, convolutional neural networks (CNNs)~\cite{wang2020multi, wang2023bugpre}, and graph neural networks (GNNs)~\cite{zhang2023detecting, cheng2021deepwukong, cao2021bgnn4vd}, deep learning has enabled the automation of tasks like bug detection, code optimization~\cite{gao2024search, cummins2020programl, baghdadi2021deep}, and code generation~\cite{le2022coderl, sethi2018dlpaper2code}. These models excel at identifying patterns within large codebases and can detect subtle issues that traditional tools may overlook. However, the application of deep learning to concurrent programming is still evolving, as the unique intricacies of concurrency require specialized datasets and architectures.

Recent breakthroughs in natural language processing (NLP), particularly through transformer-based large language models (LLMs) such as GPT-4, have demonstrated unprecedented capabilities in understanding and generating code. These models excel in various programming-related tasks, including code completion~\cite{eghbali2024hallucinator, ugare2024improving, liu2024stall+, chen2021evaluating}, bug detection~\cite{lu2024grace, li2024llm}, and automated code repair~\cite{alrashedy2023can, hidvegi2024cigar, jin2023inferfix, bouzenia2024repairagent, joshi2023repair} by leveraging vast training data and sophisticated neural network architectures. However, LLM-generated code can also introduce vulnerabilities due to their reliance on probabilistic patterns and potential exposure to insecure or flawed examples during training~\cite{panichella2024vulnerabilities}. Given their success and limitations in these areas, it is imperative to explore the potential of LLMs in verifying concurrent programs.

LLMs have transformed software development by offering human-like code understanding and generation. Trained on vast code and natural language datasets, these models can translate high-level descriptions into functional code, detect vulnerabilities, and aid debugging. They are particularly promising for concurrent programming, potentially identifying complex synchronization issues that rule-based tools often miss. By bridging human intent and machine execution, LLMs have become invaluable for writing and analyzing code.

Despite their strengths in code completion, bug detection, and automated generation, LLMs' capabilities in verifying concurrent programs remain underexplored. Given the complexity of concurrency issues—such as managing shared resources, synchronization primitives, and subtle bugs under relaxed memory models—evaluating how well LLMs can assist in verifying these programs is crucial. Although LLMs produce syntactically correct code, they may generate logically flawed or unsafe code~\cite{liu2024your}, a concern that is magnified in concurrent programming due to the intricate nature of potential errors.

While there have been significant advancements in concurrency analysis tools~\cite{jain2022bird, norris2013cdschecker, demsky2015satcheck}, including static analyzers, model checkers~\cite{abdulla2017stateless, kokologiannakis2020hmc, gadelha2018esbmc}, and dynamic testing frameworks, these tools typically require intricate knowledge of the underlying memory models. They can be limited in their ability to scale to large, real-world software systems. Moreover, they may not generalize well to novel concurrency patterns or emerging programming paradigms. 
LLMs, such as GPT-4 and its successors, have demonstrated remarkable natural language understanding, code generation~\cite{austin2021program, jain2022jigsaw}, and general pattern recognition capabilities. These 
models have been trained on vast amounts of textual and code data, enabling them to perform tasks ranging from code completion to bug detection. However, their potential in understanding and analyzing concurrent programs, especially under relaxed memory models, remains an open question.

As hardware systems increasingly adopt relaxed memory models to boost performance and efficiency, the need for effective verification tools has become more urgent. Understanding these models is crucial to avoid subtle bugs, such as race conditions, which can lead to unpredictable behavior, data corruption, or security vulnerabilities. Additionally, because various architectures implement different memory models, developers must understand the specifics of each model to write efficient and reliable code.

LLMs have shown impressive general-purpose reasoning abilities~\cite{liu2024codemind, cai2024codegraph}, but their proficiency in specialized and complex domains, such as concurrent programs under relaxed memory models (e.g., Total Store Order (TSO) and Partial Store Order (PSO)), remains largely untested. Assessing their reasoning capacity in these specific technical areas presents significant challenges. Unlike traditional program analysis tools, designed with a deep understanding of computational rules, LLMs lack a built-in conceptual framework for handling the low-level details of memory consistency, thread interactions, and synchronization mechanisms. This gap can lead to difficulties in accurately predicting or explaining complex behaviors in concurrent programs, making it challenging for LLMs to offer reliable guidance on memory model-related issues. Furthermore, the probabilistic nature of LLMs may produce outputs that sound plausible but are technically incorrect when dealing with the stringent and precise requirements of memory models in software development.

This paper seeks to explore the capabilities of LLMs in detecting concurrency issues within programs, with a particular focus on how well they handle the complexities introduced by relaxed memory models. Specifically, we investigate the performance of LLMs in reasoning about TSO and PSO, two widely studied relaxed memory models critical for understanding modern concurrency. These models are chosen because TSO allows for some instruction reordering while maintaining a relatively strong consistency guarantee, and PSO introduces further flexibility by permitting even more relaxed ordering of store operations. By analyzing these two models, we aim to answer three primary research questions through experiments and case studies.

\begin{itemize}
    \item \textbf{RQ1:} How accurately can LLMs understand and summarize concurrent programs\textsuperscript{$\dagger$}?
    \item \textbf{RQ2:} How can LLMs effectively identify common concurrency issues such as race conditions, deadlocks, and synchronization errors in concurrent programs?
    \item \textbf{RQ3:} How do LLMs compare to traditional static and dynamic analysis tools in verifying concurrent programs in terms of both accuracy and efficiency?
    \item \textbf{RQ4:} Can LLMs predict correct program behaviors under relaxed memory models?
\end{itemize}

\footnotetext{$\dagger$ Unless explicitly stated, \textit{concurrent programs} refer to\textit{ concurrent programs under a sequentially consistent memory model} in this paper.}

By investigating these research questions, we aim to provide a comprehensive evaluation of the capabilities of LLMs in the context of concurrent programming. 
Understanding these aspects is crucial for integrating LLMs into software development workflows, potentially enhancing tools for debugging and verifying concurrent code. The findings of this study could pave the way for more advanced and efficient tools for verifying concurrent programs, ultimately contributing to the development of more reliable and robust software systems.

The main contributions of this paper are as follows.

\begin{enumerate}
    \item \textit{Prompt-Based Evaluation:} The paper introduces a set of targeted prompts designed to assess LLMs on their understanding of concurrency issues in programs. These prompts focus on various aspects of concurrent programming, such as thread management and synchronization.

    \item \textit{Performance Insights}: The paper provides a detailed analysis of LLM performance in understanding and summarizing concurrency-related problems by applying these prompts. This analysis highlights the strengths and limitations of LLMs in this context.

    \item \textit{Benchmarking LLMs}: The study sets benchmarks for evaluating LLMs based on their prompt responses. This approach facilitates a structured assessment of LLM capabilities and provides a basis for future evaluations.
\end{enumerate}
Overall, the paper advances the understanding of LLMs in concurrent program analysis by evaluating their performance through specific questions, aiming to enhance the integration of LLMs into software development practices. To promote transparency and facilitate further research, the artifacts associated with this work, including the experimental datasets, code samples, and LLM responses, are publicly available at \url{https://anonymous.4open.science/r/LLM-Conc-RMM-13F8}.

The structure of this paper is as follows: Section~\ref{related} provides an overview of prior research and related work. Section~\ref{sec:method} details our approach to evaluating the LLMs. The experimental setup and results are presented in Section~\ref{sec:experiment}, which also addresses the research questions posed in Section~\ref{introduction}. Section~\ref{limitations} discusses the findings, while Section~\ref{future} explores potential future directions and opportunities. Finally, Section~\ref{conclusion} summarizes and concludes this work.
\section{Related Work}
\label{related}

The field of concurrency analysis and verification has evolved over the years, with various methodologies being applied to identify and address the challenges of analyzing multi-threaded programs. This section reviews the most relevant related work.

\subsection{Traditional Verification Tools}
Traditional verification tools, including both static~\cite{jourdan2016verasco, gadelha2018esbmc, beyer2018cpa, wolf2021gobra, bu2022brick} and dynamic analysis~\cite{jain2022bird, singh2021dynamic, mutlu2014dynamic, bao2016polycheck, khanna2018dynamic}, have long been employed to identify concurrency issues in concurrent programs. Static analysis tools analyze code without executing it, relying on heuristics and symbolic execution to detect potential issues such as race conditions and deadlocks~\cite{blackshear2018racerd, costea2023hippodrome}. These tools are often highly effective in finding certain errors, but they can struggle with complex, dynamic interactions in concurrent programs. They may produce many false positives~\cite{johnson2013don}, where potential issues are flagged but would not occur in actual execution, and false negatives~\cite{cui2024empiricalstudyfalsenegatives}, where subtle concurrency bugs go undetected~\cite{kester2010good}.

In contrast, dynamic analysis tools monitor the execution of a program to detect concurrency issues in real time~\cite{gu2018dynamic, thokair2023dynamic, xu2020krace}. These tools typically have a lower false-positive rate in detecting real-world concurrency problems since they observe the actual execution behavior. However, their effectiveness is limited by the coverage of the test cases used during execution~\cite{thokair2023dynamic}. Dynamic analysis can be computationally expensive and may introduce significant overhead, slowing program execution. Moreover, such tools might miss issues manifest only under specific conditions or rarely encountered inputs. The balance between accuracy and efficiency and the challenge of minimizing false positives and negatives highlights the complexity and limitations inherent in traditional verification approaches for concurrent programs.

\subsection{Deep Learning for Code Analysis}
Deep learning (DL) models, particularly those trained on large-scale code repositories, have shown significant potential in capturing intricate patterns and semantics within code, enabling advanced tasks such as code completion~\cite{hu2019deep, svyatkovskiy2019pythia}, bug detection, and vulnerability identification. Integrating these techniques into code analysis has facilitated the development of more sophisticated models that can learn from vast amounts of data, especially from open-source software (OSS) communities.

 Traditional code analysis techniques often rely on rule-based systems~\cite{havelund2015rule, de2015rule} or symbolic reasoning~\cite{ramos2015under, cadar2021klee, bliudze2015formal}, which may struggle with the complexity and variability of real-world code. In contrast, deep learning models, particularly those based on neural networks, have demonstrated the ability to learn patterns and representations from large code data~\cite{carpenter1989neural, hellendoorn2017deep}. With increasing accuracy, these models can be trained to perform tasks such as code completion, bug detection~\cite{li2017software, phan2017convolutional}, and vulnerability identification~\cite{cao2021bgnn4vd, wang2023deepvd}. Recent developments in neural code representation, such as Graph Neural Networks (GNNs) and Transformer-based architectures, have further enhanced the capability to capture the intricate dependencies and control flows in code~\cite{leclair2020improved, cheng2021deepwukong, zhou2019devign}.  
 These methods have shown to be useful in identifying data races~\cite{zhang2022novel, tehrani2019deeprace}, deadlocks~\cite{hasanzad2013prediction}, and other concurrency-related issues~\cite{qin2024cross}. 
 However, while deep learning models have succeeded in various code analysis tasks, their performance in handling concurrency, especially under relaxed memory models, remains challenging. This gap underscores the need for further research, particularly in the context of LLMs, which have begun to show potential in analyzing and summarizing concurrent code. Integrating deep learning techniques with LLMs presents a promising direction for advancing the state of the art in concurrent code analysis.

\subsection{Large Language Models in Code Analysis}
The rapid advancements in LLMs have enabled their application across various domains, including code-related tasks~\cite{nichols2023modeling, jain2022jigsaw}. These models have shown promising results in enhancing programming error messages, making them more interpretable and actionable for novice programmers~\cite{leinonen2023using}.
One of the key areas where LLMs have been employed is content analysis(~\cite{nicholas2023lost}. LLMs have also shown to enhance the decompilation process by translating binary code into high-level representations~\cite{hu2024degpt, tan2024llm4decompile}. Leveraging their training on both source code and binaries, they can improve decompilers by providing insights. As these models become increasingly integrated into technical systems, it is crucial to understand their limitations and potential pitfalls, particularly in the context of code content analysis.

LLMs are trained on vast source code and natural language datasets, enabling them to identify patterns indicative of bugs~\cite{jin2023inferfix, li2024enhancing}, such as common coding errors, logical inconsistencies~\cite{sosareasoning, zhang2024detecting}, or security vulnerabilities~\cite{ferrag2023securefalcon, ashiwal2024llm}. In bug detection, LLMs can analyze code snippets to predict potential issues~\cite{deng2023large, yang2024large}~\cite{eberhardt2024vulngpt}, often providing explanations or suggestions for correction. LLMs can automatically generate patches for bug repair by rewriting faulty code or proposing changes that align with best practices~\cite{pearce2023examining, wei2023copiloting}~\cite{xia2023automated, joshi2023repair, bouzenia2024repairagent, compcode}. However, patches generated by LLMs may not always be trustworthy~\cite{llmAPRuntrust, pearce2022examining, evalplus, tian2023assistant, liu2024codeevaluation}.

Recent research suggests LLMs can enhance various stages of the testing lifecycle~\cite{wang2024software, kande2024security}. They automate test case generation from natural language requirements and code specifications, reducing manual effort and improving test coverage. In CI/CD pipelines, LLMs generate and update test scripts, ensuring alignment with ongoing code ~\cite{zhang2024effectiveness}. They also contribute to code coverage analysis by identifying untested regions and suggesting targeted tests, while in bug detection and repair, they identify programming errors and propose fixes. LLMs enhance automated code reviews by enforcing coding standards and identifying issues pre-integration, and they prioritize regression testing based on recent code changes. Furthermore, LLMs are used in fuzzing~\cite{oliinyk2024fuzzing, xia2024fuzz4all, yang2023kernelgpt}, where they generate diverse and unexpected input data to uncover edge cases and vulnerabilities, expanding the scope of traditional testing methods. 

LLMs can be integrated with static analysis tools to assist in the automated program verification~\cite{wen2024enchanting, charalambous2023new}, bug detection and repair~\cite{jin2023inferfix, berabi2024deepcode}. These tools leverage the LLMs' ability to understand and generate code to suggest fixes for concurrency-related bugs or to verify the correctness of synchronization mechanisms. However, while LLMs show promise in catching common and well-understood bugs, they still face challenges in addressing complex, domain-specific issues or understanding the full context of the software~\cite{ullah2024llms}.

\subsection{Gaps and Challenges}

Despite the rapid advancements in LLMs for code analysis, applying these models to concurrent programs remains under-explored~\cite{nichols2024can,  mahmud2023autoparllm, valero2023comparing, schneider2024mpirigen, chen2023data}. While LLMs have shown proficiency in various programming tasks, such as code generation, bug detection, and refactoring, their ability to understand and analyze concurrency—a critical aspect of modern software—has not been extensively studied. Concurrent programs introduce complexities such as thread interactions, synchronization issues, and race conditions, which are challenging for even the most advanced models to interpret and reason about accurately. Moreover, the situation becomes even more intricate when considering relaxed memory models, like TSO and PSO, where traditional assumptions about memory consistency do not hold. These models allow for non-intuitive instruction reordering, making the correct program behavior analysis particularly difficult. The lack of research in this area highlights a significant gap in the current capabilities of LLMs, underscoring the need for further investigation into how these models can be adapted or enhanced to effectively handle the unique challenges posed by concurrent programming under relaxed memory models.

\section{Methodology}
\label{sec:method}
\begin{figure}[t]
    \centering
    \includegraphics[width=1\textwidth]{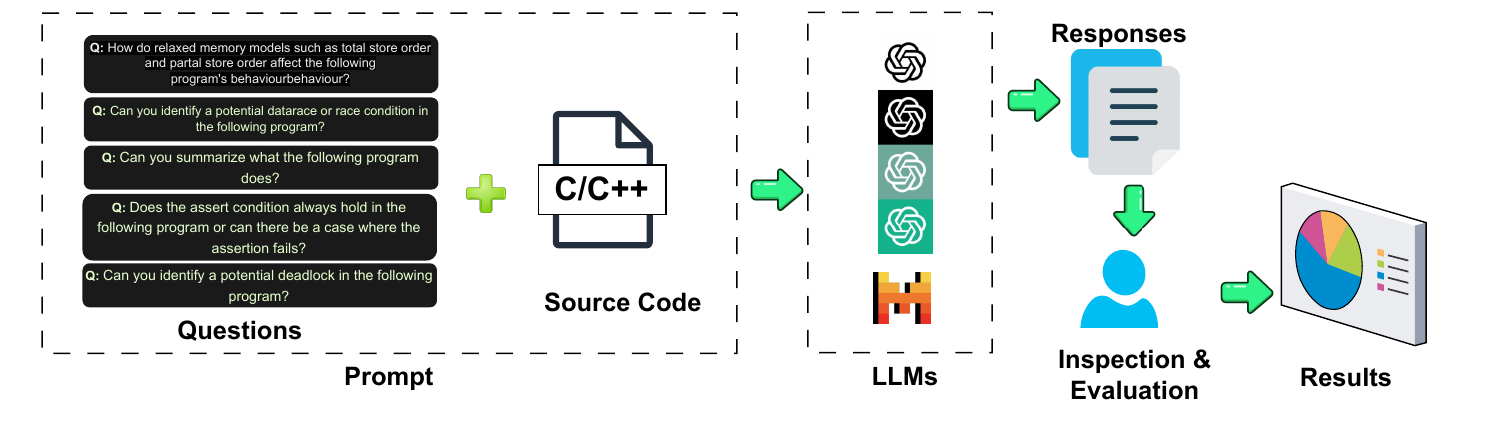}
    \caption{An overview of our approach.}
    \label{fig:overview}
\end{figure}

This section outlines the approach used to evaluate the capabilities of LLMs in verifying concurrent programs under relaxed memory models. Our approach is divided into several phases, as shown in Figure~\ref{fig:overview}, to assess these capabilities systematically, each focusing on a critical aspect of the analysis process.

\begin{table*}[t]
    \centering
    \scriptsize
    \renewcommand{\arraystretch}{1}
    \rowcolors{2}{gray!25}{white}
\begin{tabular}{|p{2cm}|p{2cm}|p{2.5cm}|p{2.5cm}|p{2.5cm}|p{2.5cm}|}
\hline
\textbf{Feature} & \textbf{GPT-4o} & \textbf{GPT-4} & \textbf{GPT-3.5-turbo} & \textbf{GPT-40-mini}  & \textbf{Mistral-Large2}\\ 
\hline
\textbf{Initial Release Date} & May 13, 2024 & March 14, 2023 & Nov, 2022 & July 24, 2024 & July 24, 2024\\ 
\hline
\textbf{Modality}\footnotemark[2] &  Text and Images & Text and Images & Text and Images & Text and Images & Text and Images\\ 
\hline

\textbf{Output Tokens}\footnotemark[2]  & 4096 & 8192 & 4096 & 16384 & 8192 \\ 
\hline
\textbf{Parameters}~\cite{xu2024survey, schreiner2023,brown2020language} & 70B\footnotemark[1] & 1.8T\footnotemark[3] & 175B & 8B & 123B\\ 
\hline
\textbf{Cost per 1M token}~\cite{openai2024, sheremetov2023} & Input: \$5 & Input: \$30 & Input: \$3 & Input: \$0.15 & Input: \$3\\ 
& Output: \$15 & Output: \$60 & Output: \$6 & Output: \$0.6 & Output: \$9\\ 
\hline
\textbf{Context window}\footnotemark[4]~\cite{xu2024survey} & 128k & 8192 & 16385 & 128k & 128k \\ 

\hline

\textbf{Trained on Data Until} & Oct' 23 & Sept' 21 & Sept' 21 & Oct' 23 & Unknown\\ 

\hline

\end{tabular}
\caption{Details of the LLMs selected for the assessment.}
\label{tab:models}
\end{table*}

\begin{table}[t]
    \centering

        \begin{tabular}{|p{4cm}|p{9.5cm}p{1cm}|}
        \cline{1-3}
        \multirow{1}{*}{\makecell{\textbf{Program Comprehension}}} & 
        \begin{enumerate}[label=\textbf{Q\arabic*.}]
            \item Can you summarize what the following program does?
            
        \end{enumerate} 
        & \multirow{1}{0.8\textwidth}{$\left.\rule{0pt}{1\normalbaselineskip}\right\}\text{\textbf{RQ1}}$} \\
        \cline{1-1}
        
        \multirow{3}{*}{\makecell{\textbf{Concurrency Issues} \\\textbf{Identification}}} & 
        \begin{enumerate}[label=\textbf{Q\arabic*.},resume, start=2]
            \item Can you identify a potential data race in the following program?
            \item Can you identify a potential deadlock in the following program?
            \
        \end{enumerate} 
        & \multirow{3}{0.8\textwidth}{$\left.\rule{0pt}{2\normalbaselineskip}\right\}\text{\textbf{RQ2 }}$} \\
       
        \cline{1-1}
        
         \multirow{3}{*}{\textbf{Verification under SC}} & 
        \begin{enumerate}[label=\textbf{Q\arabic*.},resume, start=4]
            \item Does the assert condition always hold in the following program or can there be a case where the assertion fails?
            
        \end{enumerate} 
        & \multirow{3}{0.8\textwidth}{$\left.\rule{0pt}{1\normalbaselineskip}\right\}\text{\textbf{RQ3 }}$} \\
         \cline{1-1}    

        \cline{1-1}
        
         \multirow{3}{*}{\textbf{Verification under RMM}} & 
        \begin{enumerate}[label=\textbf{Q\arabic*.},resume, start=5]
            \item How do relaxed memory models such as total and partial store orders affect the following program's behavior?
            
        \end{enumerate} 
        & \multirow{3}{0.8\textwidth}{$\left.\rule{0pt}{1\normalbaselineskip}\right\}\text{\textbf{RQ4 }}$} \\
         \cline{1-1}

        \hline
    \end{tabular}
        \caption{Grouped Evaluation Questions}
    \label{tab:prompts}
\end{table}

\footnotetext[1]{\url{https://docs.google.com/spreadsheets/d/1kc262HZSMAWI6FVsh0zJwbB-ooYvzhCHaHcNUiA0_hY}}

\footnotetext[2]{\url{https://platform.openai.com/docs/models/gpt-4-turbo-and-gpt-4}}

\footnotetext[3]{\url{https://explodingtopics.com/blog/gpt-parameters}}

\footnotetext[3]{\url{https://platform.openai.com/docs/models/gpt-4o}}

\subsection{Program Selection}

In this study, we aim to assess the comprehension of LLMs in analyzing concurrent programs under Sequential Consistency (SC) as well as two widely studied relaxed memory models: Total Store Order (TSO) and Partial Store Order (PSO). SC is the most intuitive memory model where operations from all threads are executed in a global order that respects program order within each thread. TSO is a relaxed memory model that allows the reordering of load instructions before store instructions, as long as these events are independent. On the other hand, PSO further relaxes by allowing both loads and stores to be reordered. Unlike TSO, which ensures that stores are propagated in the order they appear within a thread, PSO allows independent stores to be propagated out of order. This introduces additional flexibility and potential performance gains but also increases the complexity of reasoning about memory behavior. By examining LLM performance across SC, TSO, and PSO, we aim to understand how these models can reason about complex memory interactions and identify concurrency issues under different memory model constraints.

LLMs may operate better within limited context windows and token sizes (and the concurrent programs are already complex), so we carefully select small concurrent programs for our experiments. We use the \textit{pthread} benchmarks from the SV-COMP dataset to evaluate concurrency under SC, with code lengths ranging from 39 to 172 lines. We utilize specially designed litmus tests ranging from 38 to 74 lines of code to capture the behavior under these specific memory models for TSO and PSO.

\subsection{Prompt Creation}

Appropriate prompts must be passed to the LLM to generate relevant results. Thus, we carefully craft questions that are passed along with the code to the LLM. These prompts are listed in Table~\ref{tab:prompts}. We categorize these prompts into four categories:
\begin{enumerate}
    \item \textbf{Program Comprehension:} LLMs show promise in 
   LLMs show promise in understanding and summarizing code. The prompts in this category are designed to assess the LLM's ability to understand the functionality and intent of the program. These questions ask the LLM to describe the purpose of the code, explain how the threads interact, and identify key variables and operations. By evaluating how well the LLMs can comprehend multi-threaded programs, we aim to gauge their general understanding of concurrency constructs like threads, atomic operations, and synchronization mechanisms.
    \item \textbf{Concurrency Issues Identification:} One of the primary objectives is to assess whether LLMs can correctly identify concurrency-related issues such as data races, deadlocks, and race conditions. LLMs can parse intricate code structures, identify parallelism, and understand synchronization constructs like mutexes, semaphores, and atomic operations by leveraging extensive training on diverse codebases. The prompts in this category focus on asking the LLM to pinpoint areas in the code where these issues might arise. For instance, questions may prompt the LLM to check for shared variable accesses without proper synchronization or potential deadlock situations due to improper locking mechanisms. This category aims to evaluate the LLM's ability to detect concurrency bugs in both simple and complex scenarios.

    \item \textbf{Verification under Sequential Consistency (SC):} Sequential consistency is a simpler and more intuitive memory model where operations across threads are executed globally. In this category, the prompts evaluate whether the LLM can verify program behavior under SC, where issues such as data races may still arise despite the memory model's simplicity. These prompts ask the LLM to reason about thread execution order and whether assertions or invariants in the code can be violated. This category is essential for benchmarking LLMs' performance under the most straightforward concurrency scenarios.
    
    \item \textbf{Verification under Relaxed Memory Models (RMM):} Relaxed memory models such as TSO and PSO introduce complexity in concurrent program behavior, as instructions can be reordered, potentially leading to unexpected outcomes. The prompts in this category are designed to test the LLM's understanding of program behavior under such memory models. For example, questions may ask the LLM whether the assertion in a program will always hold under TSO or PSO or if the reordering of instructions might lead to subtle bugs. This set of prompts helps us assess the LLMs' capability to reason about non-intuitive behaviors introduced by relaxed memory models.

\end{enumerate}

    \subsection{Prompting LLMs}

    The selection of LLMs for this study was driven by the need to evaluate models with diverse capabilities in understanding and reasoning about concurrent programs. We chose a range of state-of-the-art models, including GPT-4, GPT-3.5-turbo, GPT-4o, GPT-4o-mini, and Mistral-AI's Large2, for their varying architectural strengths, token handling capacities, and performance across natural language processing tasks, as shown in Table~\ref{tab:models}. These models represent different levels of reasoning complexity and context comprehension, which is critical for assessing how well they can handle the intricacies of concurrency and memory models.

    Effective interaction with LLMs requires carefully crafted prompts that elicit relevant and precise responses. Given the intricacies of concurrency and memory model behavior, we designed targeted prompts that guide the LLMs toward identifying key aspects of the programs. Each LLM was given a task framed as a question or instruction. Since LLMs can produce verbose output, we designed prompts encouraging focused and succinct responses.

\subsection{Manual Inspection}
LLMs often generate verbose responses with portions that may be correct, incorrect, or irrelevant to the task, so a thorough manual examination of each output is essential. After prompting the LLMs, we carefully reviewed every response to assess its factual accuracy, relevance to the question asked, and the level of detail provided.

During this manual labeling process, special attention is paid to whether the LLMs correctly handle more subtle issues, such as race conditions or memory reordering under TSO and PSO models. We also consider whether the responses provide feasible error traces or if the LLMs merely speculate about potential issues without justifying their conclusions.

Once categorized, the responses are labeled accordingly, with correct information highlighted as useful and reliable, while incorrect or irrelevant segments are flagged for further analysis. This systematic evaluation ensures a comprehensive understanding of each LLM’s ability to handle complex concurrency scenarios, enabling us to identify areas where the models excel or struggle.

\section{Evaluation}
\label{sec:experiment}
\subsection{Experimental Setup and Objectives}
We conducted a series of experiments using a well-defined setup to evaluate the capabilities of LLMs in verifying concurrent programs. This section details the hardware and software environments, datasets used, and procedures followed to ensure the reproducibility and robustness of our results. Our experiments were performed on an Intel x86/64 Linux machine with Ubuntu 20.04, 32GB RAM, an Intel Core i7-4510U CPU with a clock speed of up to 3.10GHz, and kernel version 5.14.0-1029-oem. Notably, these experiments were conducted using pre-trained LLMs without any fine-tuning.

\subsection{Dataset Description}

The dataset combines 25 ARM Litmus tests~\cite{alglave2011litmus} designed for relaxed memory models, specifically TSO and PSO, with the \textit{pthread} benchmarks from the Software Verification Competition (SV-COMP)~\cite{beyer2023sv-comp}. The Litmus tests examine behaviors such as message passing, store buffering, load buffering, read consistency, and write-read ordering to identify and analyze potential anomalies in memory ordering and consistency under different relaxed memory models. The \textit{pthread} benchmarks cover a variety of concurrency scenarios, including simple locking mechanisms, the producer-consumer problem, dining philosophers, reader-writer locks, and barrier synchronization. These benchmarks evaluate multi-threaded programs' correctness and safety properties using \textsc{pthreads}. Together, these datasets provide comprehensive coverage for studying concurrency and memory model behaviors, offering theoretical analysis through \textit{litmus} tests and practical evaluation through \textit{pthread} benchmarks. Note that all selected tests contain information on our interest in concurrency issues. We consider this information as \textit{ground truth}. However, we strip these tests of any comments they contain to eliminate biases.

\subsection{Model Selection}

Selecting an appropriate model is a critical step in evaluating the capabilities of lLLMs for verifying concurrent programs. This study considers well-established LLMs such as Mistral-AI, GPT-3.5-turbo, and GPT-4, which are known for their superior language understanding and generation capabilities. These models have demonstrated state-of-the-art performance in various natural language processing tasks, making them prime candidates for our investigation. Additionally, we will compare the performance of these models against smaller, more specialized models to assess whether larger models' increased complexity and resource demands translate into significantly better performance. 

In this study, we have chosen a range of models, including GPT-3.5-turbo, GPT-4, GPT-4-o, GPT-4o-mini, and Mistral Large-latest to cover a spectrum of capabilities, from advanced general-purpose language processing to specialized tasks. GPT-3.5-turbo was selected for its efficiency and balance between performance and computational cost, making it ideal for scenarios where rapid prototyping and cost-effective solutions are needed. GPT-4, known for its enhanced reasoning abilities and broader context handling, serves as our benchmark for complex language understanding tasks. The GPT-4-o and GPT-4o-mini variants are included to explore trade-offs between performance and model size, allowing us to evaluate the scalability and adaptability of the GPT-4 architecture. Finally, Mistral's Latest, a cutting-edge model known for its specialization in certain domains, is selected to assess its ability to handle tasks that require domain-specific knowledge or enhanced generative capabilities. 

The temperature setting for all models was maintained at their respective default values. The temperature parameter controls the randomness of the model's output, with lower values resulting in more deterministic responses and higher values producing more creative or diverse outputs. Using the default temperature settings ensures a fair comparison across models, allowing each to operate within its optimal configuration for generating coherent and contextually appropriate responses.

\begin{table}[t]
    \centering
    \scriptsize
    \renewcommand{\arraystretch}{1}
    \rowcolors{2}{gray!25}{white}
\begin{tabular}{|ll|p{0.2cm}cc|ccc|ccc|ccc|ccc|ccc|}
\hline
\multicolumn{2}{|c|}{} & \multicolumn{3}{c|}{\textbf{Ground Truth}} & \multicolumn{3}{c|}{\makecell{\textbf{Mistral-AI}\\\textbf{Large2}}} & \multicolumn{3}{c|}{\textbf{GPT-3.5-turbo}} & \multicolumn{3}{c|}{\textbf{GPT-4o-mini}}  & \multicolumn{3}{c|}{\textbf{GPT-4}} & \multicolumn{3}{c|}{\textbf{GPT-4o}} \\
\cline{1-20}
Test  & LoC & Q2 & Q3 & Q4& Q2 & Q3 & Q4 & Q2 & Q3 & Q4 & Q2 & Q3 & Q4 & Q2 & Q3 & Q4 &  Q2 & Q3 & Q4   \\
\hline
2.2W & 44 & \xmark & \xmark & \xmark & \xmark & \xmark & \cmark & \cmark & \cmark & \cmark & \xmark & \xmark & \xmark & \xmark & \xmark & \cmark & \xmark & \xmark & \xmark   \\

CoWR & 41 & \xmark & \xmark & \xmark & \cmark & \xmark & \cmark & \cmark & \cmark & \cmark & \cmark & \xmark & \cmark & \xmark & \xmark & \cmark & \xmark & \xmark & \xmark   \\

RWC+dmbs & 54 & \xmark & \xmark & \xmark & \cmark & \xmark & \cmark & \cmark & \cmark & \cmark & \cmark & \cmark & \xmark & \xmark & \xmark & \xmark & \xmark & \xmark & \xmark  \\

R+dmbs &43 & \xmark & \xmark & \xmark & \cmark & \xmark & \cmark & \cmark & \cmark & \cmark & \cmark & \xmark & \cmark & \cmark & \xmark & \cmark & \cmark & \xmark  & \xmark \\

R & 44 & \xmark & \xmark & \xmark & \cmark & \xmark & \cmark & \cmark & \cmark & \cmark & \cmark & \xmark & \cmark & \cmark & \xmark & \cmark & \cmark & \xmark & \cmark  \\

SB+dmbs& 63 & \xmark & \xmark & \xmark& \xmark & \xmark & \cmark & \cmark & \cmark & \cmark & \cmark & \xmark & \cmark & \xmark & \xmark & \xmark & \xmark & \xmark & \xmark  \\

SB & 58 & \xmark & \xmark & \xmark & \xmark & \xmark & \cmark & \cmark & \cmark & \cmark & \cmark & \cmark & \cmark & \xmark & \cmark & \cmark & \cmark & \xmark & \xmark  \\

R+po+dmb & 43 & \xmark & \xmark & \xmark & \cmark & \xmark & \cmark & \cmark & \cmark & \cmark & \xmark & \xmark & \xmark & \xmark & \xmark & \cmark & \xmark & \xmark & \cmark  \\

R+dmb+po & 47 & \xmark & \xmark & \xmark & \cmark & \xmark & \cmark & \cmark & \cmark & \cmark & \cmark & \xmark & \cmark & \cmark & \xmark & \cmark & \cmark & \xmark  & \xmark \\

WRW+2W+dmbs & 60 & \xmark & \xmark & \xmark & \cmark & \xmark & \cmark & \cmark & \cmark & \cmark & \cmark & \xmark & \cmark & \cmark & \xmark & \cmark & \xmark & \xmark & \cmark  \\

RWC+dmb+po &53 &\xmark & \xmark & \xmark & \cmark & \xmark & \cmark & \cmark & \cmark & \cmark & \cmark & \xmark & \cmark & \xmark & \xmark & \cmark & \xmark & \xmark & \xmark  \\

CoRR1 &40 & \xmark & \xmark & \xmark & \cmark & \xmark & \cmark & \cmark & \cmark & \cmark & \cmark & \xmark & \cmark & \xmark & \xmark & \xmark & \xmark & \xmark & \cmark \\

2+2W+reads & 62 & \xmark & \xmark & \xmark & \cmark & \xmark & \cmark & \cmark & \cmark & \cmark & \xmark & \xmark & \cmark & \cmark & \xmark & \cmark & \xmark & \xmark & \xmark \\

2+2W+dmbs & 39 & \xmark & \xmark & \xmark & \xmark & \xmark & \cmark & \cmark & \cmark & \cmark & \xmark & \xmark & \cmark & \xmark & \xmark & \cmark & \xmark & \xmark & \xmark \\

WRW+2W & 59& \xmark & \xmark & \xmark & \xmark & \xmark & \cmark & \cmark & \cmark & \cmark & \xmark & \xmark & \cmark & \cmark & \xmark & \cmark & \xmark & \xmark & \xmark 
\\
S+po+dmb &52 & \xmark & \xmark & \xmark & \cmark & \xmark & \cmark & \cmark & \cmark & \cmark & \xmark & \xmark & \cmark & \xmark & \xmark & \cmark & \xmark & \xmark & \xmark  \\

CoRW &38& \xmark & \xmark & \xmark & \cmark & \xmark & \cmark & \cmark & \cmark & \cmark & \cmark & \xmark & \cmark & \xmark & \xmark & \cmark & \xmark & \xmark & \xmark \\

CoRR2&74 & \xmark & \xmark & \xmark & \cmark & \xmark & \cmark & \cmark & \cmark & \cmark & \cmark & \xmark & \cmark & \cmark & \xmark & \cmark & \xmark & \xmark & \xmark \\

S+dmbs &53& \xmark & \xmark & \xmark & \xmark & \xmark & \cmark & \cmark & \cmark & \cmark & \cmark & \xmark & \cmark & \xmark & \xmark & \cmark & \cmark & \xmark & \cmark \\

S &52& \xmark & \xmark & \xmark & \cmark & \xmark & \cmark & \cmark & \cmark & \cmark & \cmark & \xmark & \cmark & \cmark & \xmark & \cmark & \xmark & \xmark & \xmark  \\

2+2W+dmbs+reads &63& \xmark & \xmark & \xmark & \cmark & \xmark & \cmark & \cmark & \cmark & \cmark & \cmark & \xmark & \cmark & \cmark & \xmark & \cmark & \xmark & \xmark & \xmark  \\

RWC &57& \xmark & \xmark & \xmark & \cmark & \xmark & \cmark & \cmark & \cmark & \cmark & \cmark & \xmark & \xmark & \xmark & \xmark & \cmark & \xmark & \xmark & \xmark \\

2+2W+dmbs+po &39& \xmark & \xmark & \xmark & \xmark & \xmark & \cmark & \cmark & \cmark & \cmark & \xmark & \xmark & \cmark & \cmark & \xmark & \cmark & \xmark & \xmark & \xmark  \\

SB+dmb+po &54& \xmark & \xmark & \xmark & \cmark & \xmark & \cmark & \cmark & \cmark & \cmark & \cmark & \xmark & \xmark & \xmark & \xmark & \cmark & \xmark & \xmark & \xmark  \\

WRW+2W+dmb+po &57& \xmark & \xmark & \xmark& \xmark & \xmark & \cmark & \cmark & \cmark & \cmark & \xmark & \xmark & \xmark & \cmark & \xmark & \cmark & \xmark & \xmark & \xmark  \\

\hline
\hline
\textbf{True Positive (TP)} & & & & & 0 & 0 & 0 & 0 & 0 & 0 & 0 & 0 & 0 & 0 & 0 & 0 & 0 & 0 & 0  \\

\textbf{False Positive (FP)} && &  & & 17 & 0 & 24 & 25 & 25 & 25 & 16 & 2 & 20 & 11 & 0 & 23 & 3 & 0 & 5  \\

\textbf{True Negative (TN)} &&  & & & 8 & 25 & 1 & 0 & 0 & 0 & 9 & 23 & 5 & 14 & 25 & 3 & 22 & 25 & 20  \\

\textbf{False Negative (FN)} & & & & & 0 & 0 & 0 & 0 & 0 & 0 & 0 & 0 & 0 & 0 & 0 & 0 & 0 & 0 & 0  \\
\hline
\hline
\textbf{Accuracy} & & & & & 0.32 & \textbf{1.00} & 0.04 & 0 & 0 & 0 & 0.36 & 0.92 & 0.20 & 0.56 & \textbf{1.00} & 0.12 & \textbf{0.88} & \textbf{1.00} & \textbf{0.80}  \\
\hline
\end{tabular}   
\caption{Comparison of different LLMs' ability to detect concurrency issues in multithreaded programs from litmus tests.}
    \label{tab:rq2-litmus}
\end{table}



\begin{table}[t]
    \centering
    \scriptsize
    \renewcommand{\arraystretch}{1}
    \rowcolors{2}{gray!25}{white}
\begin{tabular}{|ll|p{0.23cm}p{0.23cm}p{0.23cm}|p{0.26cm}p{0.26cm}c|p{0.26cm}cp{0.26cm}|p{0.26cm}p{0.26cm}p{0.26cm}|p{0.25cm}cp{0.26cm}|p{0.26cm}cp{0.26cm}|}
\hline
\multicolumn{2}{|c|}{} & \multicolumn{3}{c|}{\textbf{Ground Truth}} & \multicolumn{3}{c|}{\makecell{\textbf{Mistral-AI}\\\textbf{Large2}}} & \multicolumn{3}{c|}{\textbf{GPT-3.5-turbo}} & \multicolumn{3}{c|}{\textbf{GPT-4o-mini}}  & \multicolumn{3}{c|}{\textbf{GPT-4}} & \multicolumn{3}{c|}{\textbf{GPT-4o}} \\
\cline{1-20}
Test &LoC  & Q2 & Q3 & Q4& Q2 & Q3 & Q4 & Q2 & Q3 & Q4 & Q2 & Q3 & Q4 & Q2 & Q3 & Q4 &  Q2 & Q3 & Q4   \\
\hline

triangular-2 &53& \xmark & \xmark & \cmark & \cmark & \xmark & \cmark & \cmark & \xmark & \cmark & \cmark & \cmark & \xmark & \cmark & \xmark & \cmark & \cmark & \cmark & \xmark   \\

singletonwith-unint-problems-b  &60& \cmark & \xmark & \cmark & \cmark & \xmark & \cmark & \cmark & \cmark & \cmark & \cmark & \xmark & \xmark & \cmark & \xmark & \cmark & \cmark & \xmark & \cmark   \\

stateful-01-1 &63 & \xmark & \xmark & \cmark & \xmark & \xmark & \xmark & \cmark & \cmark & \cmark & \cmark & \cmark & \xmark & \xmark & \xmark & \cmark & \cmark & \cmark & \xmark   \\

stack-2 &131 & \xmark & \xmark & \cmark & \cmark  & \cmark & \cmark & \cmark & \cmark & \cmark & \cmark & \xmark & \xmark & \xmark & \xmark & \cmark & \cmark & \cmark & \cmark   \\

triangular-longest-2 &53 & \xmark & \xmark & \cmark & \cmark & \xmark & \cmark & \cmark & \cmark & \cmark & \cmark & \cmark & \cmark & \xmark & \xmark & \cmark & \cmark & \cmark & \cmark   \\

bigshot-s& 40& \xmark & \xmark & \xmark & \xmark & \xmark & \xmark & \cmark & \cmark & \cmark & \cmark & \xmark & \xmark & \cmark & \xmark & \xmark & \cmark & \xmark & \xmark   \\

stack-longer-1 &127& \xmark & \xmark & \cmark & \cmark & \cmark & \cmark & \cmark & \cmark & \cmark & \cmark & \cmark & \cmark & \cmark & \xmark & ! & \cmark & \cmark & \cmark   \\

two-stage-3  &131& \xmark & \xmark & \cmark & \xmark & \xmark & \cmark & \cmark & \cmark & \cmark & \cmark & \cmark & \xmark & \cmark & \xmark & \cmark & \cmark & \cmark & \cmark   \\

stateful-01-2  &63& \xmark & \xmark & \xmark & \xmark & \xmark & \xmark & \cmark & \cmark & \cmark & \cmark & \cmark & \xmark & \xmark & \xmark & \xmark & \cmark & \cmark & \xmark   \\

sync-01 &  71&\xmark & \xmark & \xmark & \xmark & \cmark & \cmark & \cmark & \cmark & \cmark & \cmark & \xmark & \cmark & \xmark & \xmark & \xmark & \cmark & \xmark & \xmark   \\

reorder-5 & 84& \cmark & \xmark & \xmark & \cmark & \xmark & \cmark & \cmark & \cmark & \cmark & \cmark & \xmark & \cmark & \cmark & \xmark & \cmark & \cmark & \xmark & \cmark   \\

bigshot-s2 & 40& \xmark & \xmark & \xmark & \xmark & \xmark & \xmark & \cmark & \cmark & \cmark & \cmark & \xmark & \xmark & \xmark & \xmark & \cmark & \cmark & \xmark & \cmark   \\

reorder-2 & 86&  \cmark & \xmark & \cmark & \cmark & \xmark & \cmark & \cmark & \xmark & \cmark & \cmark & \xmark & \cmark & \cmark & \xmark & \cmark & \cmark & \xmark & \xmark   \\

indexer &87&  \xmark & \xmark & \xmark & \xmark & \cmark & \xmark & \cmark & \cmark & \cmark & \cmark & \cmark & \xmark & \cmark & \cmark & \xmark & \cmark & \cmark & \xmark   \\

queue-ok-longer &156&  \xmark & \xmark & \xmark & \xmark & \cmark & \cmark & \cmark & \cmark & \cmark & \cmark & \cmark & \cmark & \xmark & \xmark & \cmark & \cmark & \cmark & \cmark   \\

bigshot-p &39 &  \cmark & \xmark & \cmark & \cmark & \xmark & \cmark & \cmark & \cmark & \cmark & \cmark & \xmark & \cmark & \cmark & \xmark & \cmark & \cmark & \cmark & \cmark   \\

stack-1 &128&  \xmark & \xmark & \cmark & \cmark & \xmark & \xmark & \cmark & \cmark & \cmark & \cmark & \cmark & \cmark & \xmark & \xmark & \xmark & \cmark & \cmark & \cmark   \\

queue-longer &170&  \xmark & \xmark & \cmark & \cmark & \cmark & \cmark & \cmark & \cmark & \cmark & \cmark & \cmark & \cmark & \xmark & \xmark & \cmark & \cmark & \cmark & \cmark   \\

triangular-longer-1 &52&  \xmark & \xmark & \xmark & \cmark & \xmark & \xmark & \cmark & \cmark & \cmark & \cmark & \xmark & \cmark & \xmark & \xmark & \cmark & \cmark & \xmark & \cmark   \\

queue-ok &156&  \xmark & \xmark & \xmark & \xmark & \cmark & \cmark & \cmark & \cmark & \cmark & \cmark & \cmark & \cmark & \xmark & \xmark & \cmark & \cmark & \cmark & \cmark   \\

triangular-longer-2 &52&  \xmark & \xmark & \cmark & \cmark & \xmark & \cmark & \cmark & \cmark & \cmark & \cmark & \xmark & \cmark & \xmark & \xmark & \cmark & \cmark & \xmark & \cmark   \\

queue-ok-longest & 155& \xmark & \xmark & \xmark & \xmark & \cmark & \cmark & \cmark & \cmark & \cmark & \cmark & \cmark & \cmark & \xmark & \xmark & \cmark & \cmark & \xmark & \cmark   \\

triangular-longest-1 &53&  \xmark & \xmark & \xmark & \xmark & \xmark & \cmark & \cmark & \cmark & \cmark & \cmark & \cmark & \xmark & \xmark & \xmark & \cmark & \cmark & \cmark & \xmark   \\

sigma & 55&  \cmark & \xmark & \cmark & \cmark & \xmark & \cmark & \cmark & \cmark & \cmark & \cmark & \xmark & \cmark & \cmark & \xmark & \cmark & \cmark & \cmark & \xmark   \\

singletonwith-unint-problems &66&  \cmark & \xmark & \cmark & \cmark & \xmark & \cmark & \cmark & \cmark & \cmark & \cmark & \cmark & \xmark & \cmark & \xmark & \cmark & \cmark & \xmark & \xmark   \\

singleton & 66& \cmark & \xmark & \cmark & \cmark & \xmark & \cmark & \cmark & \cmark & \cmark & \cmark & \xmark & \cmark & \cmark & \xmark & \cmark & \cmark & \xmark & \cmark   \\

singleton-b & 60& \cmark & \xmark & \cmark & \cmark & \xmark & \cmark & \cmark & \cmark & \cmark & \cmark & \xmark & \cmark & \xmark & \xmark & \cmark & \cmark & \xmark & \cmark   \\

stack-longest-2 &128 & \xmark & \xmark & \xmark & \cmark & LL & \cmark & \cmark & \cmark & \cmark & \cmark & \cmark & \cmark & \xmark & \xmark & \xmark & \cmark & \cmark & \cmark   \\

queue &172& \xmark & \xmark & \cmark & \cmark & \cmark & \cmark & \cmark & \cmark & \cmark & \cmark & \cmark & \cmark & \xmark & \cmark & \cmark & \cmark & \cmark & \cmark   \\

lazy-01 &56&  \xmark & \xmark & \cmark  & \xmark & \xmark & \cmark & \cmark & \cmark & \cmark & \cmark & \cmark & \cmark & \xmark & \xmark & \cmark & \cmark & \cmark & \cmark   \\

stack-longest-1 &129&  \xmark & \xmark & \cmark  & \cmark & \cmark & \cmark & \cmark & \cmark & \cmark & \cmark & \cmark & \cmark & \cmark & \xmark & \cmark & \cmark & \cmark & \cmark   \\

triangular &52 & \xmark & \xmark & \xmark & \cmark & \xmark & \cmark & \cmark & \cmark & \cmark & \cmark & \xmark & \cmark & \xmark & \xmark & \cmark & \cmark & \xmark & \cmark   \\

stack-longer-2 & 127 & \xmark & \xmark & \xmark & \xmark & LL & \xmark & \cmark & \cmark & \cmark & \cmark & \xmark & \xmark & \xmark & \xmark & \xmark & \cmark & \xmark & \xmark   \\

queue-longest & 170 & \xmark & \xmark & \cmark & \cmark & \cmark & \cmark & \cmark & \cmark & \cmark & \cmark & \cmark & \cmark & \cmark & \xmark & \cmark & \cmark & \cmark & \cmark   \\

\hline
\hline
\textbf{True Positive (TP)} & & & &  & 8 & 0 & 18 & 7 & 0 & 20 & 8 & 0 & 14 & 7 & 0 & 18 & 8 & 0 & 15   \\

\textbf{False Positive (FP)} & & & &  & 13 & 13 & 8 & 27 & 32 & 14 & 26 & 19 & 8 & 7 & 2 & 8 & 26 & 20 & 8  \\

\textbf{True Negative (TN)} & & & & & 13 & 21 & 6 & 0 & 2 & 0 & 0 & 15 & 6 & 19 & 32 & 6 & 0 & 14 & 6  \\

\textbf{False Negative (FN)} & & & & & 0 & 0 & 2 & 0 & 0 & 0 & 0 & 0 & 6 & 1 & 0 & 2 & 0 & 0 & 5  \\
\hline
\hline

\textbf{Precision} & & & & & 0.38 & 0 &\textbf{ 0.69} &  0.21& 0 & 0.59 &0.24& 0 & 0.64 & 0.50 & 0 & \textbf{0.69 }& 0.24 & 0 & 0.65  \\

\textbf{Recall} & & & & & \textbf{1.00} & 0 & 0.90 & \textbf{1.00} & 0 & \textbf{1.00} & \textbf{1.00} & 0 & 0.70 & 0.88 & 0 & 0.90 & \textbf{1.00} & 0 & 0.75  \\
\hline
\hline
\textbf{Accuracy} & & & & & 0.62 & 0.62 & 0.71 &  0.21 & 0.06 & 0.59 & 0.24& 0.44 & 0.59 & 0.76 & \textbf{0.94} & 0.71 & 0.24 & 0.41 & 0.62 \\

\textbf{F1-Score} & & & & & 0.55 & 0 & \textbf{0.78} & 0.34 & 0 & 0.74 & 0.38  & 0 & 0.67 & 0.64 & 0 & \textbf{0.78} & 0.38 & 0 & 0.70  \\
\hline
\end{tabular}   
\caption{Comparison of different LLMs in their ability to detect concurrency issues in multithreaded programs from SV-COMP. 
\centering\textbf{LL:} Livelock. \textbf{!}: Code has assert, but LLM does not detect it. We consider LL as \cmark~and ! as \xmark~while computing TP, FP, TN and TN.} 
    \label{tab:rq2-sv}
\end{table}

\begin{table}[t]
    \centering
    \scriptsize
    \renewcommand{\arraystretch}{1}
    \rowcolors{2}{gray!25}{white}
\begin{tabular}{|l|c|cc|cc|cc|cc|cc|}
\hline
\multicolumn{1}{|c|}{} & \multicolumn{1}{c|}{\textbf{Ground Truth}} & \multicolumn{2}{c|}{\makecell{\textbf{Mistral-AI}\\\textbf{Large2}}} & \multicolumn{2}{c|}{\textbf{GPT-3.5-turbo}} & \multicolumn{2}{c|}{\textbf{GPT-4o-mini}}  & \multicolumn{2}{c|}{\textbf{GPT-4}} & \multicolumn{2}{c|}{\textbf{GPT-4o}} \\
\cline{1-12}
Test  & Q5  & Q5 & Reason & Q5 & Reason& Q5 & Reason& Q5 & Reason& Q5 & Reason  \\
\hline
2.2W  & \cmark & \xmark & \xmark & \cmark & \xmark & \cmark & \xmark & \cmark & $\circ$ & \xmark & \xmark  \\

CoWR &  \xmark & \xmark & \cmark & \cmark & \xmark & \xmark & \xmark & \xmark & \cmark & \cmark & \xmark   \\

RWC+dmbs & \xmark & \xmark & \cmark & \cmark & \xmark & \cmark & \cmark & \xmark & \cmark & \cmark & \xmark   \\

R+dmbs & \xmark & \xmark & \cmark & \cmark & \xmark & \cmark & \xmark & \xmark & \cmark & \xmark & \cmark \\

R & \cmark & \xmark & \xmark & \cmark & \xmark & \cmark & \xmark & \cmark & \xmark & \cmark  & \xmark  \\

SB+dmbs&  \xmark & \xmark & \cmark& \cmark & \xmark & \cmark & \xmark & \cmark & \cmark & \cmark & \xmark   \\

SB & \cmark & \xmark & \xmark & \cmark & \cmark & \cmark & \xmark & \cmark & $\circ$ & \cmark & \xmark  \\

R+po+dmb &  \cmark & \xmark & \xmark & - & \xmark & \cmark & \xmark & \cmark & \xmark & \xmark & \xmark  \\

R+dmb+po & \cmark & \xmark & \xmark & - & \xmark & \cmark & \xmark & \cmark & \xmark & \cmark & \xmark  \\

WRW+2W+dmbs &  \xmark & \xmark & \cmark & \cmark & \xmark & \cmark & \xmark & \xmark & \xmark & \cmark & \xmark \\

RWC+dmb+po &\cmark & \xmark & \xmark & \cmark & \xmark & \cmark & \xmark & \cmark & \xmark & \cmark & \xmark  \\

CoRR1 & \xmark & \xmark & \cmark & \xmark & \xmark & \xmark & \xmark & \cmark & \xmark & \xmark & \cmark  \\

2+2W+reads & \cmark & \xmark & \xmark & - & \xmark & \cmark & \xmark & \cmark & \cmark & \cmark & \xmark  \\

2+2W+dmbs  & \xmark & \xmark & \cmark & \cmark & \xmark & \cmark & \xmark & \cmark & \xmark & \cmark & \xmark  \\

WRW+2W &  \cmark & \xmark & \xmark & \cmark & $\circ$ & \cmark & \xmark & \cmark & \xmark & \xmark & \xmark 
\\
S+po+dmb  & \cmark & \xmark & \xmark & \cmark & \xmark & \cmark & \xmark & \cmark & \xmark & \cmark & $\circ$ \\

CoRW &  \xmark & \xmark & \cmark & \cmark & \xmark & \xmark & \xmark & \cmark & \xmark & \cmark & \xmark\\

CoRR2 &  \xmark & \xmark & \cmark & - & \xmark & \cmark & \xmark & \cmark & \xmark & \cmark & \xmark  \\

S+dmbs & \xmark & \cmark & \xmark & - & \xmark & \cmark & \xmark & \cmark & \xmark & \xmark & \cmark \\

S & \cmark & \cmark & \xmark & \cmark & \xmark & \cmark & \xmark & \cmark & \xmark & \cmark & \xmark   \\

2+2W+dmbs+reads & \xmark & \cmark & \xmark & - & \xmark & \cmark & \xmark & \xmark & \cmark & \xmark & \cmark\\\cmark

RWC & \cmark & \xmark & \xmark & \cmark & \xmark & \cmark & \xmark & \cmark & \xmark & \cmark & \xmark  \\

2+2W+dmbs+po & \cmark & \xmark & \xmark & - & \xmark & \xmark & \xmark & \cmark & \xmark & \cmark & $\circ$ \\

SB+dmb+po & \cmark & \xmark & \xmark & - & \xmark & \cmark & \xmark & \cmark & \xmark & \cmark & \xmark   \\

WRW+2W+dmb+po & \cmark & \xmark & \xmark& \cmark & \cmark & \cmark & \xmark & \cmark & \cmark & \xmark & \xmark   \\
\hline
\hline
\textbf{True Positive (TP)} &   & 2 & & 9 &  & 13 &  & 14 &  & 10 &   \\
\textbf{False Positive (FP)} &  & 2 & & 7 &  & 8 &  & 7 &  & 7 &  \\

\textbf{True Negative (TN)} &  & 8 & & 4 & & 3 & & 4 &  & 4 &  \\

\textbf{False Negative (FN)} &  & 13 & & 5 &  & 1 &  & 0 &  & 4&   \\
\hline
\hline
\textbf{Precision} &  & 0.50 & & 0.56  &  & 0.62 &  &\textbf{ 0.67} &  & 0.59&   \\

\textbf{Recall} & & 0.13 & & 0.64  &  & 0.93 &  &\textbf{ 1.00} &  & 0.71&   \\
\hline
\hline
\textbf{Accuracy} &  & 0.40 & & 0.52 &  & 0.64 &  & \textbf{0.72} &  & 0.56&   \\

\textbf{F1-Score} &  & 0.21 & & 0.60 &  & 0.74 &  & \textbf{0.80} &  & 0.65&   \\

\hline
\end{tabular}   
\caption{Comparison of different LLMs in their ability to detect concurrency issues in multithreaded programs from litmus tests under relaxed memory models.
\centering -: The LLM response does not include assertions. $\circ$: The LLM produces partial information. We consider - as \xmark~ in computing TP, FP, TN, and FN.}
    \label{tab:rq5:litmus}
\end{table}

\subsection{Prompt Creation}

To address the research questions outlined in this paper, we developed multiple prompts specifically designed to evaluate the capabilities of LLMs. These prompts were carefully crafted to assess various aspects of LLM performance, including their ability to understand, summarize, and analyze concurrent programs. By systematically applying these prompts, we aim to gain insights into the strengths and limitations of LLMs in handling the complexities of concurrency, enabling a thorough examination of their effectiveness in this domain.

Table~\ref{tab:prompts} provides a comprehensive listing of all the prompts specifically designed for querying different memory models. In the analysis process, these prompts are systematically appended to the code being examined before being submitted to the LLM for processing. The generated responses from the LLMs are then meticulously reviewed through manual examination. Our detailed findings, based on this rigorous evaluation, are presented and discussed in the subsequent subsections.

 \subsection{Structure of the Tables}

The experimental results are presented in Tables~\ref{tab:rq2-litmus},\ref{tab:rq2-sv}, and\ref{tab:rq5:litmus}. Each table's first column lists the name of the test program, providing a reference point for the specific concurrency scenario being evaluated. The second column in Tables~\ref{tab:rq2-litmus} and~\ref{tab:rq2-sv} indicates the number of lines of code (LoC) in each test program, offering insight into the complexity of the code analyzed.

The column titled \textit{Ground Truth} presents the actual, verified results for each question, serving as a benchmark against which the performance of various LLMs is measured. Following this, the columns labeled \textit{Mistral-AI}, \textit{GPT-3.5-turbo}, \textit{GPT-4o-mini}, \textit{GPT-4}, and \textit{GPT-4o} display the results generated by each respective LLM in response to the specific questions posed.

Within these columns, the sub-columns titled with Q<\textit{index number}> correspond to the questions listed in Table~\ref{tab:prompts}, which were used to construct the prompts passed to the LLMs. This detailed breakdown allows for a side-by-side comparison of how each LLM performed in identifying and addressing concurrency-related issues across different test cases. The structure of these tables facilitates a clear understanding of each model's strengths and weaknesses, providing valuable insights into their capabilities in handling complex concurrency scenarios.

Tables~\ref{tab:rq2-litmus} and~\ref{tab:rq2-sv} capture details present the results for Q2, Q3 and Q4 on different models. We mark \textit{yes} with \cmark and \textit{no} with \xmark. Since LLMs do not typically provide binary answers, all generated responses were manually evaluated before being assigned. For example, Q2 specifically asks about \textit{data races}, but the models may generate responses related to logical races or general race conditions. Similarly, in Q3, which queries \textit{deadlocks}, some LLMs may incorrectly respond with \textit{livelocks}. For Q4, which looks for potential \textit{assertion failures}, the models may miss key assert conditions in the code, leading to incorrect conclusions. 

Table~\ref{tab:rq5:litmus} provides a comparison of different LLMs in their ability to detect assertion failures in concurrent programs running under relaxed memory models, such as TSO and PSO. In these scenarios, LLMs not only generate a prediction of whether an assertion failure might occur but also attempt to provide reasoning for why the failure happens, often including possible execution traces. For each model, we not only assess whether it correctly identifies assertion failures but also evaluate the accuracy of the provided explanations. The reasoning behind an assertion failure is crucial because understanding the underlying cause is key to resolving the issue. For example, an LLM might correctly predict an assertion failure but offer an incorrect or incomplete explanation regarding thread interactions, memory ordering, or synchronization mechanisms.

\subsection{Results and Analysis}
Our analysis aims to answer the research questions outlined earlier in this paper, focusing on the LLMs' ability to handle complex concurrency patterns, detect bugs, and reason about relaxed memory models such as TSO and PSO.

 \subsubsection{Answering \textbf{RQ1: How accurately can LLMs understand and summarize concurrent programs?}}
\vspace{2pt}
 
The accuracy of LLMs in understanding and summarizing concurrent programs varies depending on several factors, including the complexity of the code, the structure of the code, and the memory model supported by the host architecture. However, for this particular research question, we prompt the LLM with the Q1 in Table~\ref{tab:prompts} along with the test code without specifying any memory model. We evaluate the selected models based on the following key points:

\begin{enumerate}
    \item \textbf{Basic Understanding:} All selected models could identify what the program did under a sequentially consistent memory model. Interestingly, while the other models produced detailed output by identifying global variables and explaining other program details such as function definitions and the assertion conditions, GPT-3.5-turbo generally captures the high-level details only.
    
    \item \textbf{Identifying Concurrency Issues:} LLMs show varying degrees of success in identifying common concurrency issues such as data races, deadlocks, and atomicity violations. While they can sometimes point out potential issues, they may miss subtleties, especially in more complex scenarios involving intricate inter-thread interactions or relaxed memory models. As per the results of our experiments reported in Table~\ref{tab:rq2-litmus}, GPT-4 outperforms all other models in identifying concurrency issues.
    
    \item \textbf{Summarization:} LLMs can generate coherent summaries of concurrent programs, highlighting key components like synchronization mechanisms and shared resources. However, the accuracy and completeness of these summaries can be inconsistent. For more complex concurrency patterns, the summaries may omit critical details or oversimplify important aspects of the program's behavior.

\end{enumerate}

 \subsubsection{Answering \textbf{RQ2:  How can LLMs effectively identify common concurrency issues such as race conditions, deadlocks, and synchronization errors in concurrent programs?}\\}

\vspace{2pt}

To address this research question, we utilized prompts Q2, Q3, and Q4 from Table~\ref{tab:prompts}, appending them to the test code for each of the selected litmus tests and SV-COMP \textit{pthread} tests. The outcomes of our experiments are systematically documented in Tables~\ref{tab:rq2-litmus} and \ref{tab:rq2-sv}, which present the results for the litmus tests and the \textit{pthread} tests, respectively. Each model's response was manually validated to ensure accuracy and reliability. These tables provide a detailed analysis of the models' performance in responding to the prompts across different test scenarios, offering insights into their ability to understand and process the complexities of concurrent programming under various conditions.

Prompts Q2 and Q3 are curated to evaluate LLMs' capabilities in identifying data races and deadlocks, respectively. Prompt Q4 is designed to investigate whether LLMs can detect assertion failures caused by thread interactions in concurrent programs. We use atomic load and store instructions with default sequentially consistent ordering to evaluate LLM's understanding of the program behavior under a sequentially consistent memory model. 

Prompts Q2 and Q3 were specifically curated to evaluate the capabilities of LLMs in identifying data races and deadlocks, respectively. In contrast, prompt Q4 is designed to probe whether LLMs can detect assertion failures that arise from thread interactions in concurrent programs. We employed atomic load and store instructions to assess the LLMs' understanding of program behavior under a sequentially consistent memory model, ensuring they followed the default sequentially consistent ordering. This approach allows us to rigorously test the models' ability to accurately interpret and analyze the behavior of concurrent programs in environments where memory operations are expected to maintain a strict order.

As shown in Tables~\ref{tab:rq2-litmus} ans~\ref{tab:rq2-sv}, Mistral-AI's Large2, GPT-4, and GPT-4o outperform the other models. The results vary slightly between the Litmus tests and SV-COMP’s \textit{pthread} tests. For instance, in the Litmus tests, Mistral-AI's Large2 achieves the highest accuracy in detecting deadlocks, while in the SV-COMP tests, GPT-4 performs best in this category.

It is important to note that in Table~\ref{tab:rq2-litmus}, none of the tests exhibit data races, deadlocks, or assertion failures under the SC memory model, as they are specifically designed to trigger bugs under the relaxed memory models, TSO and PSO. Consequently, for all tests, the True Positives and False Negatives are zero, leading to Precision, Recall, and F1-Score values of zero. As a result, we report only the Accuracy for these cases.

 \subsubsection{Answering \textbf{RQ3:  How do LLMs compare to traditional static and dynamic analysis tools in verifying concurrent programs in terms of both accuracy and efficiency?}\\}

\vspace{2pt}

Some LLMs can effectively identify common concurrency issues such as race conditions, deadlocks, and synchronization errors in concurrent programs by leveraging their ability to analyze code, understand control flow, and recognize patterns associated with concurrency problems. 
The tests listed in Table~\ref{tab:rq2-litmus} are manually curated from the ARM Litmus Tests~\cite{alglave2011litmus}, with all load and store operations on global variables defined as atomic. Consequently, these tests are free of data races. Additionally, they have been carefully designed to avoid any deadlock scenarios. Assertion failures in these tests were verified using the Nidhugg tool, with the sequentially consistent memory model specified by the \textsc{--sc} option.
The ground truth for the results reported in Table~\ref{tab:rq2-sv} was generated using the popular bounded model checking tool ESBMC~\cite{gadelha2018esbmc}. This ground truth was further manually verified (with the \textsc{.yml} files provided with the tests) to ensure accuracy, providing a reliable baseline against which the performance of the LLMs can be evaluated.

We consider the following key criteria to evaluate LLMs in their ability to identify common concurrency issues.

\begin{enumerate}
    \item \textbf{Race condition}
    \begin{itemize}
        \item \textit{Pattern Recognition:}
        LLMs can be trained to recognize common code patterns that lead to race conditions, such as multiple threads accessing shared variables without proper synchronization. By analyzing how variables are read, written, and modified in concurrent code, LLMs can identify scenarios where interleaving operations may lead to inconsistent states. Our experiments show that while LLMs have limitations in accurately predicting the absence of data races, GPT-4o performed the best among the selected LLMs.
        \item \textit{Semantic Understanding:}
        Contextual Understanding: LLMs can analyze the interactions between different threads and the resources they access. They can evaluate whether appropriate locks or synchronization mechanisms, such as mutexes, semaphores, or atomic operations, are employed. LLMs can flag potential issues by identifying instances where shared data is accessed without adequate protection. In our experiments, all selected LLMs demonstrated the ability to correctly identify synchronization primitives.

    \end{itemize}
    \item \textbf{Deadlock}
    \begin{itemize}
        \item \textit{Locking Order Analysis:}
        LLMs can analyze the sequence in which locks are acquired in a program, helping to identify potential deadlocks. Deadlocks often arise when threads attempt to acquire multiple locks in an inconsistent order. By examining the sequence of lock acquisitions across threads, LLMs can detect circular dependencies or problematic lock hierarchies that might lead to deadlocks. Our experiments reveal that while LLMs accurately identified locks and synchronization mechanisms, some models incorrectly detected deadlocks. However, LLMs such as Mistral-AI, GPT-4, and GPT-4o achieved a 100 percent accuracy rate for litmus tests.
        \item \textit{State Space Exploration:}LLMs can simulate or reason about various execution paths in concurrent programs, exploring potential states where deadlocks might occur. LLMs can identify and highlight possible deadlock scenarios by simulating thread interactions and resource acquisitions before they manifest. Our observations show that all selected models, except GPT-3.5-turbo, generated feasible execution traces for the test program.
             
    \end{itemize}

\end{enumerate}

Furthermore, Table~\ref{tab:rq5:litmus} presents the results of the LLMs' capabilities in identifying assertion failures that may occur under relaxed memory models, specifically TSO and PSO. The ground truth for these tests was determined using the Nidhugg tool with the memory model option \textsc{--pso}, and these results were also manually verified for correctness. This thorough verification process ensures that the LLMs' outputs can be accurately assessed against established standards, allowing for a robust evaluation of their ability to handle complex concurrency issues under various memory models."

\subsubsection{Answering \textbf{RQ4: Can LLMs predict correct program behaviors under relaxed memory models?}\\}

The 25 litmus tests selected for this study were specifically chosen to address Research Question 5 (RQ5), which examines the capabilities of LLMs in understanding program behavior under relaxed memory models. These tests are uniquely tailored to assess the complexities of relaxed memory models, particularly TSO and PSO. To ensure that the LLMs accurately capture the semantics of relaxed memory, all read and write operations are specified as atomic and tagged with \textit{std::memory\_order\_relaxed}. Table~\ref{tab:rq5:litmus} provides key insights in response to Research Question 5 (RQ5), offering a comparative analysis of the models evaluated:

\begin{enumerate}
    \item \textbf{Performance in Identifying Relaxed Memory Model Behavior:} Among the evaluated models, GPT-4 demonstrates superior performance in identifying behaviors associated with relaxed memory models, such as TSO and PSO. However, despite its relative strength, GPT-4's performance remains inferior to specialized existing tools designed for this specific purpose.

    \item \textbf{Accuracy in Identifying Assertion Failures:} Models like GPT-4 and GPT-4o show a commendable ability to accurately identify assertion failures triggered under relaxed memory models, specifically TSO and PSO. Nevertheless, these models often fail to provide correct explanations for the causes of these assertion failures. While they generally describe the intra-thread reordering allowed by TSO and PSO accurately, they incorrectly attribute the assertion failures to inter-thread reordering when applying their reasoning to the test program.

    \item \textbf{Handling of Memory Fences:} LLMs including, GPT-4, GPT-4o and Mistral-Large2 correctly recognize scenarios where memory fences constrain instruction reordering. In these cases, the models accurately determine that the assertion holds, acknowledging the role of memory fences in preserving program order and ensuring correctness. 

    \item \textbf{Interleaving Generation:} While nearly all evaluated models—except GPT-3.5-turbo—were able to generate some feasible traces, none successfully captured the intra-thread reorderings permitted by the relaxed memory models. Despite this shortcoming, the detailed outputs provided by the models do offer explanations of how events can be reordered within a thread. Interestingly, in some instances, the models mistakenly included load-load reorderings under Partial Store Order (PSO), which are, in fact, restricted by this memory model. This indicates a partial understanding but also highlights limitations in the models' comprehension of the specific constraints imposed by relaxed memory models.

\end{enumerate}

While GPT-4 and other advanced LLMs show some ability to understand relaxed memory models and identify behaviors like TSO and PSO, their performance remains noticeably weaker than that of specialized tools designed for these specific tasks. Although these models can occasionally identify assertion failures and recognize the role of memory fences in maintaining program order, they frequently fail to provide accurate explanations for the underlying causes of these failures, often incorrectly attributing them to inter-thread reorderings. Furthermore, the models struggle with correctly handling intra-thread reorderings and misunderstand restricted behaviors, such as load-load reordering under PSO, which exposes significant gaps in their understanding of the complexities inherent in relaxed memory models.

Tables~\ref{tab:rq2-litmus} and~\ref{tab:rq2-sv} provide a comparative analysis of the efficacy of different LLMs in identifying concurrency-related issues in multi-threaded programs, benchmarked against the established ground truth. These tables showcase the performance of the LLMs in detecting critical issues such as data races, deadlocks, and assertion failures, highlighting the strengths and limitations of each model.

It is important to note that the prompts passed to the LLMs during these evaluations did not explicitly specify any particular memory model. This omission intentionally assessed the LLMs' inherent ability to reason about concurrency issues without relying on predefined memory model constraints. The results captured in these tables offer insights into how well the LLMs can generalize and apply their understanding of concurrency across different scenarios, even when the underlying memory model is not explicitly defined. This approach allows us to gauge the models' adaptability and robustness in handling complex, real-world concurrency challenges.


\subsection{Case Study}
Listings~\ref{lst:corw} and~\ref{lst:22w} showcase two sample codes from the Litmus tests, CoRW and 2+2W, respectively. We provide the corresponding question posed to the LLMs and the test code for each listing. Additionally, on the same page, responses from various LLMs are presented alongside the listings for comparison.

Listing~\ref{lst:corw} presents a code snippet that creates two threads, contains no data races, and where the assertion always holds under SC, TSO, and PSO memory models. In addition, the code snippet contains atomic operations with default SC semantics. We showcase the responses generated by various LLMs when asked whether they could identify any \textit{data race} in the provided code. As LLM responses are often verbose, we include pruned versions that capture the key points. Responses are color-coded: 1) green for factually correct and relevant answers, 2) yellow for correct but non-essential or partial information, and 3) red for incorrect information.

We observe that GPT-4 generates the most factually correct and relevant information compared to the evaluated models. Although GPT-4 answers the question correctly, it occasionally includes irrelevant or incorrect details. In contrast, GPT-3.5-turbo, GPT-4o-mini, and Mistral-AI's Large2 incorrectly predict the presence of a data race in the code. GPT-4o, while producing a significant amount of irrelevant information, ultimately identifies the absence of data races correctly with default SC semantics. We showcase the responses generated by various LLMs when asked \textit{"Can you identify a potential data race or race
condition in the following program?"} along with the test program.

\begin{tcolorbox}[sidebyside, sidebyside align=top seam, width=\textwidth, boxrule=0.0pt, colback=white, colframe=white, lefthand width=0.5\textwidth, righthand width=0.45\textwidth]
\begin{lstlisting}[style=mystyle,label={lst:corw}, caption=A concurrent code example (CoRW) from the limus tests.]
#include <pthread.h>
#include <atomic>
#include <assert.h>

using namespace std;

atomic_int x;
atomic_int a;

void *thread1(void *threadid)
{
    int p;
    atomic_store(&x, 1);
    p = atomic_load(&x);
    if (p == 2)
    {
        atomic_store(&a, 1);
    }
}

void *thread2(void *threadid)
{
    atomic_store(&x, 2);
}

int main()
{
  int i=0;
  int j=1;
  int rc1,rc2;
  pthread_t threads[2];
  rc1 = pthread_create(&threads[0], NULL,
                          thread1, (void *)i);
  rc2 = pthread_create(&threads[1], NULL, 
                          thread2, (void *)j);
  (void) pthread_join(threads[0], NULL);
  (void) pthread_join(threads[1], NULL);
  assert( x != 1 || a != 1);
}



\end{lstlisting}

\vspace{10pt}

\textbf{\textit{Q}} \textit{Can you identify a potential data race or race condition in the following program?}\\
\begin{tcolorbox}
    
\begin{minipage}[t]{0.3\textwidth} 
    \includegraphics[width=0.5cm,height=0.5cm]{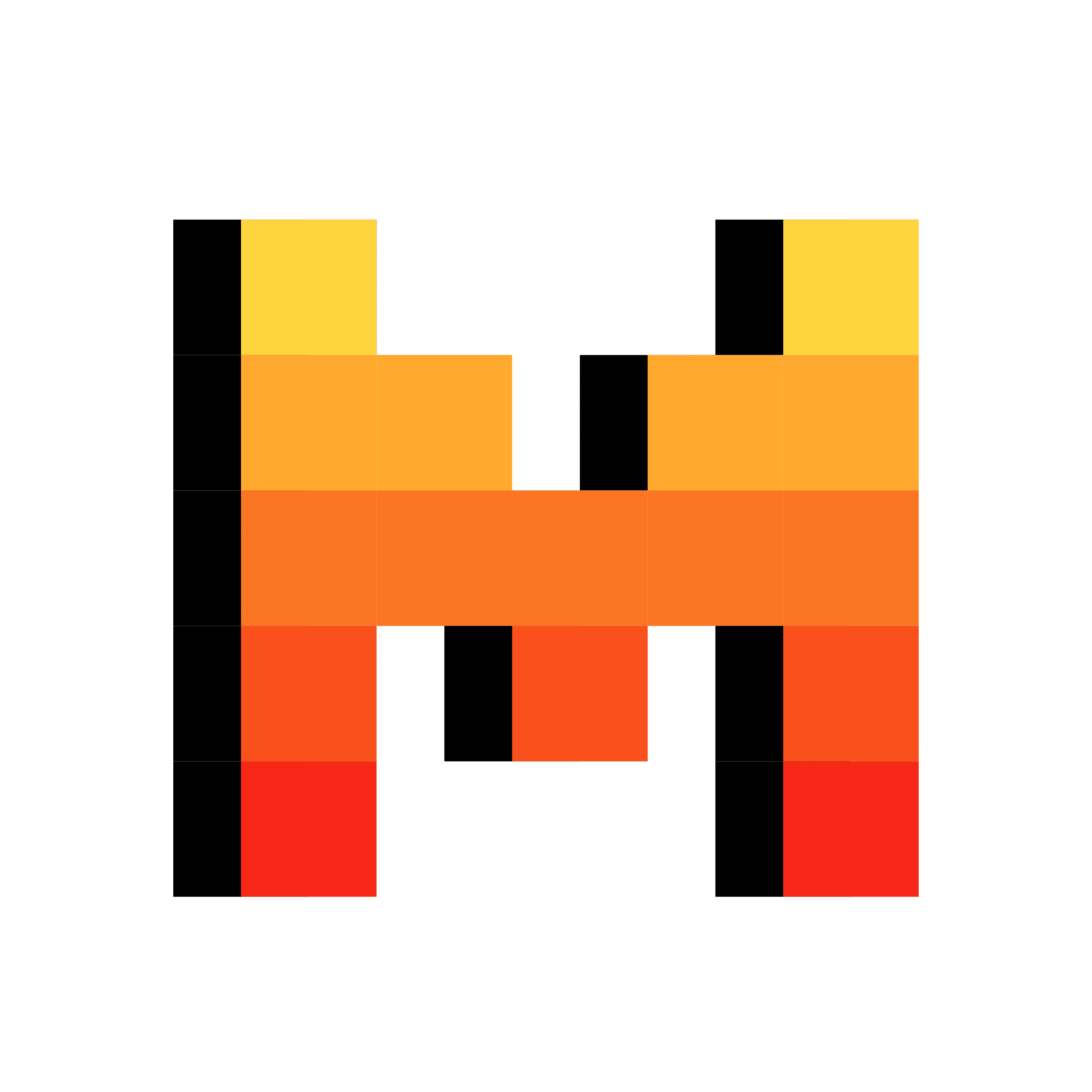}
\end{minipage}%
\begin{minipage}[t]{0.65\textwidth} 
    
    \scriptsize\textbf{Mistral-AI Response}
\end{minipage}

\scriptsize
\textcolor{lightred}{
Yes, there is a potential data race in the given program. The data race occurs because the two threads can access and modify the shared atomic variables `x` and `a` concurrently without proper synchronization, leading to unpredictable behavior.
}
\textcolor{darkgreen}{
The race condition arises because the order of execution of these operations is not deterministic. Here are some possible scenarios:
}
\\
\textit{Scenario 1:}
\begin{itemize}
    \item  \textcolor{darkyellow}{Thread 1 stores `1` into `x`.}
    \item  \textcolor{darkyellow}{Thread 2 stores `2` into `x`.}
    \item  \textcolor{darkyellow}{Thread 1 loads `2` from `x` into `p`.}
    \item  \textcolor{darkyellow}{Since `p` is `2`, Thread 1 stores `1` into `a`.}
\end{itemize}

\textit{Scenario 2:}
\begin{itemize}
    \item \textcolor{darkyellow}{ Thread 1 stores `1` into `x`.}
    \item \textcolor{darkyellow}{ Thread 1 loads `1` from `x` into `p`.}
    \item \textcolor{darkyellow}{ Since `p` is not `2`, Thread 1 does not store `1` into `a`.}
    \item \textcolor{darkyellow}{Thread 2 stores `2` into `x`.}
\end{itemize}

\textcolor{lightred}{In the first scenario, both `x` and `a` could end up being `1`, which would violate the assertion `assert(x != 1 || a != 1)`.}

\end{tcolorbox}

\begin{minipage}{0.3\textwidth}
    \centering
    \begin{tcolorbox}[colback=darkgreen,boxrule=0mm,width=0.15cm,height=0.15cm,boxsep=0mm]
    \end{tcolorbox}
    \vspace{0.1cm}
    \textbf{True \& Relevant}
\end{minipage}%
\begin{minipage}{0.3\textwidth}
    \centering
    \begin{tcolorbox}[colback=darkyellow,boxrule=0mm,width=0.15cm,height=0.15cm,boxsep=0mm]
    \end{tcolorbox}
    \vspace{0.1cm}
    \textbf{True but Irrelevant}
\end{minipage}%
\begin{minipage}{0.3\textwidth}
    \centering
    \begin{tcolorbox}[colback=lightred,boxrule=0mm,width=0.15cm,height=0.15cm,boxsep=0mm]
    \end{tcolorbox}
    \vspace{0.1cm}
    \textbf{False Statement}
\end{minipage}

\tcblower

\section*{}

\vspace{-10pt}

\begin{tcolorbox}

\begin{minipage}[t]{0.3\textwidth} 
    \includegraphics[width=0.3cm,height=0.3cm]{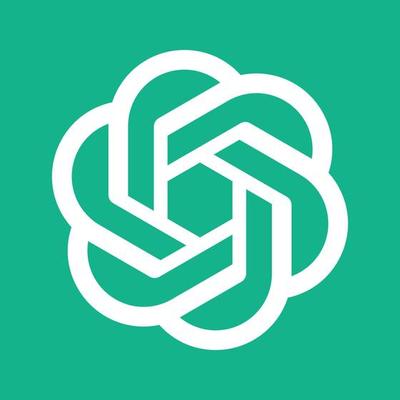}
\end{minipage}%
\begin{minipage}[t]{0.65\textwidth} 
    
    \scriptsize\textbf{GPT-3.5-turbo Response}
\end{minipage}
\scriptsize
\textcolor{lightred}{
Yes, there is a potential data race in the program. The data race occurs when both `thread1` and `thread2` access and modify the shared atomic variable `x` without proper synchronization. 
}

\textcolor{darkgreen}{In `thread1` function, `x` is first loaded and then compared to 2 to store a value in the atomic variable `a`.} \textcolor{lightred}{However, in `thread2` function, `x` is directly modified without any synchronization. If `thread2` modifies `x` after `thread1` has loaded but before the comparison is made in `thread1`, then the assertion at the end of `main` may fail.}

 \textcolor{darkyellow}{To resolve this potential data race, you can use proper synchronization mechanisms such as mutex locks or atomic operations to ensure that only one thread can access and modify the shared variable `x` at a time.}
\end{tcolorbox}
\vspace{-10pt}

\begin{tcolorbox}

\begin{minipage}[t]{0.3\textwidth} 
    \includegraphics[width=0.3cm,height=0.3cm]{images/gpt.jpg}
\end{minipage}%
\begin{minipage}[t]{0.65\textwidth} 
    
    \scriptsize\textbf{GPT-4 Response}
\end{minipage}
\scriptsize

\textcolor{darkgreen}{No, there is no data race in this program. All the shared data – `x` and `a` – are accessed using atomic operations, guaranteeing that the operations are performed in a thread-safe manner. There is, however, a race condition between `thread1` and `thread2` regarding the value of `x`.}

\textcolor{lightred}{ In `thread1`, if `x` equals `2`, it will set `a` to `1`. In `thread2`, `x` is set to `2`. However, the order in which thread1 and thread2 are scheduled to run is not deterministic. Therefore, it's possible `thread1` executes before `thread2` and hence the if-condition `p == 2` will never be true and `a` would never be set to `1`. On the other hand, if `thread2` runs before `thread1`, `x` is set to `2` and thus `a` will be set to `1`. These represent two different execution outcomes based on scheduling, which is a race condition (related to timing and sequencing), but not a data race.}

\end{tcolorbox}
\vspace{-10pt}

\begin{tcolorbox}

\begin{minipage}[t]{0.3\textwidth} 
    \includegraphics[width=0.3cm,height=0.3cm]{images/gpt.jpg}
\end{minipage}%
\begin{minipage}[t]{0.65\textwidth} 
    
    \scriptsize\textbf{GPT-4o Response}
\end{minipage}
\scriptsize

\textcolor{darkyellow}{To identify a potential data race in the given program, let's analyze the interplay between the two threads and the shared atomic variables `x` and `a`.}

\textcolor{darkyellow}{Here's a breakdown of the execution flow:}
\begin{itemize}
\item \textcolor{darkyellow}{Thread 1 (`thread1` function):}
   - \textcolor{darkyellow}{Loads the value of `x` into `p`.}
   - \textcolor{darkyellow}{Stores `1` into `x`.}
   - \textcolor{darkyellow}{If `p` is `2`, then stores `1` into `a`.}

\item \textcolor{darkyellow}{Thread 2 (`thread2` function):}
   - \textcolor{darkyellow}{Stores `2` into `x`.}

\item \textcolor{darkyellow}{Main function:}
   - \textcolor{darkyellow}{Creates two threads running `thread1` and `thread2`.}
   - \textcolor{darkyellow}{Joins the threads.}
   - \textcolor{darkyellow}{Asserts that `a != 1 || x != 2`.}
\end{itemize}

\textcolor{darkyellow}{In the context of atomic operations:}
\begin{itemize}
\item \textcolor{darkyellow}{Atomicity and Atomic Operations:}
   -\textcolor{darkgreen}{ Atomic operations themselves do not have data races because they complete without interference.}
   -\textcolor{darkyellow}{ However, the program's logic could still lead to unintended outcomes, depending on the order of atomic operations between the two threads.}
   \end{itemize}
\end{tcolorbox}
\vspace{-10pt}

\begin{tcolorbox}

\begin{minipage}[t]{0.3\textwidth} 
    \includegraphics[width=0.3cm,height=0.3cm]{images/gpt.jpg}
\end{minipage}%
\begin{minipage}[t]{0.65\textwidth} 
    
    \scriptsize\textbf{GPT-4o-mini Response}
\end{minipage}
\scriptsize

\textcolor{lightred}{The program you provided has potential data races due to improper synchronization between the threads. We need to analyze the code step by step to identify potential issues.}

\textbf{ Potential Data Race Analysis}
\textcolor{darkyellow}{\textit{Thread Scheduling:}}
\begin{itemize}
    \item \textcolor{darkyellow}{The execution of thread 1 and thread 2 is non-deterministic. The order in which they run can vary.}
    \item \textcolor{darkyellow}{If thread 1 runs first, it will store `1` in `x` and subsequently load the value of `x`. If thread 2 has not yet executed, `p` will be `1`, and `a` will not be set.}
    \item \textcolor{darkyellow}{If thread 1 runs first, it will store `1` in `x` and subsequently load the value of `x`.}
    \item \textcolor{darkyellow}{If thread 2 runs first, it sets `x` to `2` before thread 1 reads the value. In this case, when thread 1 reads `x`, it finds `p == 2`, and thus `a` is set to `1`.}
\end{itemize}

\end{tcolorbox}

\vspace{-10pt}

\end{tcolorbox}

\begin{tcolorbox}[sidebyside, sidebyside align=top seam, width=\textwidth, boxrule=0.5pt, colback=white, colframe=white, lefthand width=0.5\textwidth, righthand width=0.45\textwidth]
\begin{lstlisting}[style=mystyle, label={lst:22w},caption=A concurrent code example (2+2W) from the limus tests.]
#include <iostream>
#include <pthread.h>
#include <atomic>
#include <assert.h>

std::atomic<int> x;
std::atomic<int> y;

void *thread1(void *threadid)
{
  
    x.store(2, std::memory_order_relaxed);
    y.store(1, std::memory_order_relaxed);
}

void *thread2(void *threadid)
{
    y.store(3, std::memory_order_relaxed);
    x.store(4, std::memory_order_relaxed);
}

int main()
{
  int i=0;
  int j=1;
  int rc1,rc2;
  pthread_t threads[2];
  rc1 = pthread_create(&threads[0], NULL,
                          thread1, (void *)i);
  rc2 = pthread_create(&threads[1], NULL, 
                          thread2, (void *)j);
  (void) pthread_join(threads[0], NULL);
  (void) pthread_join(threads[1], NULL);

  assert( x != 2 || y != 3 );

}




\end{lstlisting}
\vspace{10pt}

\textbf{\textit{Q}} \textit{How do relaxed memory models such as total store order and partial store order affect the following program's behavior?}\\

\begin{tcolorbox}
    
\begin{minipage}[t]{0.3\textwidth} 
    \includegraphics[width=0.5cm,height=0.5cm]{images/mistral.png}
\end{minipage}%
\begin{minipage}[t]{0.65\textwidth} 
    
    \scriptsize\textbf{Mistral-AI Response}
\end{minipage}
\scriptsize

\textcolor{darkgreen}{Relaxed memory models, such as Total Store Order (TSO) and Partial Store Order (PSO), allow for certain reorderings of memory operations that can affect the behavior of concurrent programs. Let's analyze how these memory models might affect the given program.}

\textbf{***Generates SC Traces***}

\textcolor{lightred}{
\textbf{Assertion Analysis\\}
The assertion assert(x != 2 || y != 3) checks that it is not possible for x to be 2 and y to be 3 at the same time.\\}

\textcolor{lightred}{Given the possible outcomes listed above, it is clear that the assertion should never fail because:}
\begin{itemize}[left=0pt]
    \item \textcolor{lightred}{If x is 2, then y must be 1 (since Thread 1 sets y to 1 after setting x to 2).}
    \item \textcolor{lightred}{If y is 3, then x must be 4 (since Thread 2 sets x to 4 after setting y to 3).}
\end{itemize}

\textcolor{lightred}{Therefore, the assertion assert(x != 2 || y != 3) will always hold true, regardless of the interleaving of the threads under TSO or PSO.}
\end{tcolorbox}


\begin{minipage}{0.3\textwidth}
    \centering
    \begin{tcolorbox}[colback=darkgreen,boxrule=0mm,width=0.15cm,height=0.15cm,boxsep=0mm]
    \end{tcolorbox}
    \vspace{0.1cm}
    \textbf{True \& Relevant}
\end{minipage}%
\begin{minipage}{0.3\textwidth}
    \centering
    \begin{tcolorbox}[colback=darkyellow,boxrule=0mm,width=0.15cm,height=0.15cm,boxsep=0mm]
    \end{tcolorbox}
    \vspace{0.1cm}
    \textbf{True but Irrelevant}
\end{minipage}%
\begin{minipage}{0.3\textwidth}
    \centering
    \begin{tcolorbox}[colback=lightred,boxrule=0mm,width=0.15cm,height=0.15cm,boxsep=0mm]
    \end{tcolorbox}
    \vspace{0.1cm}
    \textbf{False Statement}
\end{minipage}

\tcblower

\begin{tcolorbox}

\begin{minipage}[t]{0.3\textwidth} 
    \includegraphics[width=0.3cm,height=0.3cm]{images/gpt.jpg}
\end{minipage}%
\begin{minipage}[t]{0.65\textwidth} 
    
    \scriptsize\textbf{GPT-3.5-turbo Response}
\end{minipage}
\scriptsize

\textcolor{darkyellow}{In this particular program, the use of relaxed memory order for storing and loading values in the atomic variables x, y, a, and b can lead to race conditions. For example, in thread2 and thread3, there is no synchronization mechanism to ensure proper ordering of operations on the shared variables x and y. As a result, it is possible for the atomic stores and loads to be reordered in a way that violates the intended logic of the program. }
\textcolor{lightred}{Additionally, the use of std::atomic\_thread\_fence with memory\_order\_seq\_cst in thread2 and thread3 does not guarantee the necessary synchronization between threads to prevent data races. The relaxed memory orders used in combination with the fence may lead to unexpected behavior due to lack of synchronization.}

\end{tcolorbox}

\vspace{-10pt}
\begin{tcolorbox}

\begin{minipage}[t]{0.3\textwidth} 
    \includegraphics[width=0.3cm,height=0.3cm]{images/gpt.jpg}
\end{minipage}%
\begin{minipage}[t]{0.65\textwidth} 
    
    \scriptsize\textbf{GPT-4 Response}
\end{minipage}
\scriptsize
\textcolor{darkyellow}{Here are the possible scenarios for this program:}

\textcolor{darkyellow}{1. Thread1 runs then thread2 runs:
   The final values will be `x = 4, y = 3` and the assert will not fail.}

\textcolor{darkyellow}{2. Thread2 runs then thread1 runs:
   The final values will be `x = 2, y = 1` and the assert will not fail.}

\textcolor{darkgreen}{3. Thread1 and thread2 run concurrently and the operations interleave:
   The potential outcomes are harder to predict due to reordering allowed by the relaxed memory model. It's possible for either `x or y` or both to get either of their possible values from the threads (2 or 4 for `x`, 1 or 3 for `y`), and these assignments can happen in any order.}

\textcolor{darkgreen}{Therefore, With the relaxed memory model, it's theoretically possible to reach a state where `x = 2 and y = 3`, which would cause the assert to fail.}


\end{tcolorbox}
\vspace{-10pt}

\begin{tcolorbox}

\begin{minipage}[t]{0.3\textwidth} 
    \includegraphics[width=0.3cm,height=0.3cm]{images/gpt.jpg}
\end{minipage}%
\begin{minipage}[t]{0.65\textwidth} 
    
    \scriptsize\textbf{GPT-4o Response}
\end{minipage}
\scriptsize




\textcolor{darkgreen}{TSO is relatively strict, so while it allows some reordering, it guarantees that writes will appear in order from each thread's perspective.}




\textcolor{darkgreen}{Partial Store Order (PSO)
PSO allows even more flexibility than TSO, as stores can appear out of order even within a single thread.}




\textcolor{lightred}{ Even under PSO, the memory model allows more reordering and relaxation compared to TSO, but given the constraints of atomic variables and basic operations, the assertion `assert(x != 2 || y != 3);` will still hold true. The program's safe ordering practices due to atomic variables ensure that `x == 2` and `y == 3` are not observed together.}

\textcolor{lightred}{In conclusion, the assertion will not fail in either TSO or PSO due to the properties of atomic operations, even with `memory\_order\_relaxed`.}

\end{tcolorbox}

\vspace{-10pt}

\begin{tcolorbox}

\begin{minipage}[t]{0.3\textwidth} 
    \includegraphics[width=0.3cm,height=0.3cm]{images/gpt.jpg}
\end{minipage}%
\begin{minipage}[t]{0.65\textwidth} 
    
    \scriptsize\textbf{GPT-4o-mini Response}
\end{minipage}
\scriptsize
\begin{itemize}[left=0pt]
    \item \textcolor{darkgreen}{\textbf{Thread 1 Completes First:} Suppose Thread 1 executes first and sets `x` to `2` and then `y` to `1`. After this completes, we can have:
   - `x = 2` and `y = 1` - This would not trigger the assert.}
   \item  \textcolor{darkgreen}{ \textbf{Thread 2 Completes First"} If Thread 2 sets `y` to `3` first and then `x` to `4`, then afterwards:
   - `x = 4` and `y = 3` - This would also not trigger the assert.}
   \item   \textcolor{darkgreen}{ 3. \textbf{Interleavings: }However, because of relaxed memory (especially in PSO), Thread 1 and Thread 2 can have interleaved executions:
   - Thread 1 sets `x = 2`
   - Thread 2 sets `y = 3`
   - Thread 1 sets `y = 1`
   - Thread 2 sets `x = 4`
   In this final state, the values at the end would be:
   - `x = 4` and `y = 1` - This does not trigger the assert.}
   \item 
 \textcolor{darkgreen}{4. **Unexpected Visibility**: Yet, due to memory reordering, it's also possible (though implausible under these atomic operations) that Thread 2 could see `x = 2` and `y = 3` if operations were reordered at the store level (which might not normally happen but could theoretically if compilers optimized it in an unintended way).}

\end{itemize}

\end{tcolorbox}
\vspace{10pt}

\end{tcolorbox}

Similarly, the sample code in Listing~\ref{lst:22w} creates two threads that perform atomic operations on variables \textit{x} and \textit{y}. These events can be reordered according to the specified memory order \textsc{memory\_order\_relaxed} within the thread, potentially leading to an assertion failure. However, since the operations are atomic, no data race occurs. We explicitly specify memory order consistency when prompting the LLMs about the code's behavior under relaxed memory models, TSO, and PSO. The responses are pruned and color-coded for clarity.

Although reasoning about a concurrent program's behavior under relaxed memory models can be challenging for LLMs due to insufficient training data, GPT-4 correctly identifies the potential for an assertion failure under PSO memory ordering, though it also generates some irrelevant information. Interestingly, it does not produce the trace, which leads to the error. Mistral-AI Large 2 and GPT-4o, while accurately defining the memory orderings permitted under TSO and PSO, incorrectly respond that the assertion always holds, as they fail to provide a feasible trace leading to the error. Surprisingly, GPT-4o-mini offers only correct and relevant information but also fails to produce the failing trace.

Our evaluation of LLM responses to concurrent programming code shows varying performance. For Listing~\ref{lst:corw}, which tests for data races, GPT-4 is the most accurate, though it occasionally includes irrelevant details. GPT-3.5-turbo, GPT-4o-mini, and Mistral-AI Large2 mistakenly identify a data race, while GPT-4o provides mostly irrelevant information but correctly identifies the absence of data races. For Listing~\ref{lst:22w}, which involves atomic operations under relaxed memory models, GPT-4 accurately predicts potential assertion failures under PSO but does not provide the error trace. Mistral-AI Large 2 and GPT-4o correctly describe memory orderings but fail to detect assertion failures, and GPT-4o-mini, while accurate, also misses the error trace.


 
\section{Discussion}
\label{limitations}

In this section, we discuss the key strengths and limitations of LLMs in the context of concurrent program analysis based on the results of our study. 
\subsection{Strengths}

\subsubsection{Natural Language Reasoning and Explanation}
One of the core strengths of LLMs, particularly models like GPT-4, is their ability to reason about code and explain in natural language. This capability makes them highly accessible to developers who may not be well-versed in formal verification tools. LLMs can explain the potential risks or issues in concurrent code, such as race conditions or deadlocks, in an intuitive and human-readable manner. 

\subsubsection{Detection of Simple Concurrency Bugs}
LLMs can detect basic concurrency issues like race conditions and deadlocks in simpler, small-scale programs. In our study, GPT-4 and GPT-4o demonstrate the highest accuracy in successfully identifying common concurrency bugs that were present in the selected tests. This shows the potential of LLMs to serve as a first pass for identifying glaring concurrency bugs without the need for complex tool configurations.

\subsubsection{Flexibility Across Various Codebases} 
Unlike traditional static analysis tools, which often require configuration and may be language-specific, LLMs can analyze a wide range of codebases with minimal setup. This flexibility allows LLMs to be used in various programming languages and frameworks, providing a more universal solution.

\subsubsection{Understanding of High-Level Memory Model Concepts}
LLMs, especially GPT-4, reasonably understood high-level memory model concepts such as Total Store Order (TSO) and Partial Store Order (PSO). While their analysis was not always precise, their ability to reason about the effects of memory ordering on program behavior demonstrated promise.

\subsection{Challenges}

\subsubsection{Incomplete and Inconsistent Analysis}
Despite their strengths in detecting basic concurrency issues, LLMs are prone to incomplete and inconsistent analysis. The models may struggle to provide accurate or actionable insights in complex scenarios involving deep synchronization mechanisms or rare concurrency bugs. For example, GPT-4 correctly identified some concurrency bugs but often missed subtle issues, such as those arising from intricate interleavings of thread execution. This inconsistency limits the reliability of LLMs for comprehensive program verification.

\subsubsection{Lack of Formal Guarantees}

Unlike traditional verification tools, which can provide formal guarantees of program correctness, LLMs rely on probabilistic models that generate responses based on learned patterns. As a result, LLMs cannot offer the same level of certainty in their analysis. For instance, while models like ESBMC~\cite{gadelha2018esbmc} and Nidhugg~\cite{abdulla2017stateless} systematically explore all possible execution paths to identify concurrency issues, LLMs may overlook important edge cases or provide incorrect solutions that only appear plausible.

\subsubsection{Difficulty in Handling Relaxed Memory Models}
Although LLMs showed some understanding of relaxed memory models, they struggled to handle the intricacies of TSO and PSO effectively. The models often failed to provide detailed error traces or identify subtle reordering issues that could lead to assertion failures. 

\subsubsection{Verbose and Irrelevant Responses}
Another notable limitation of LLMs is the tendency to produce verbose responses, often including irrelevant or tangential information. In many cases, LLMs would provide lengthy explanations that mixed correct and incorrect information, making it difficult to extract useful insights. This verbosity can hinder developers who need concise, actionable feedback on concurrency issues.

\section{Future Work}
\label{future}
Our evaluation of LLMs for verifying concurrent programs under relaxed memory models highlights several exciting future research opportunities. One key direction is to enhance the integration between LLMs and traditional verification tools to improve the accuracy and depth of automated verification. Developing interfaces that allow LLMs to effectively utilize verification results could enable the automation of complex verification tasks. A promising approach is creating hybrid systems that combine the LLM’s natural language reasoning with classical verification methods, potentially allowing these systems to automatically generate and validate patches or bug fixes. 

Further, current LLMs struggle to accurately handle the subtleties of concurrency and memory models like TSO and PSO. Future research could focus on fine-tuning LLMs using datasets specifically curated for concurrent programming and memory model behaviors. By training LLMs on a broader set of code examples involving complex thread interactions, non-deterministic behaviors, and relaxed memory model effects, we can potentially improve their performance in identifying concurrency bugs and reasoning about reordering scenarios.

Another promising direction is the automatic generation of test cases for concurrency issues. LLMs could be utilized to generate synthetic test cases designed to expose subtle bugs in multi-threaded programs under various memory models.

While this study focused on TSO and PSO memory models, other relaxed memory models, such as Release Consistency (RC) and Weak Memory Consistency (WMC), are used in various hardware architectures. Future research could expand LLM analysis to include these additional memory models. Additionally, extending LLM capabilities to cover a broader range of programming languages commonly used for concurrent programming, such as Rust, Go, and Erlang, would enhance the study's relevance. Supporting multiple languages would enable wider adoption and applicability of LLMs across diverse industries and software systems, beyond the scope of the C/C++ programs evaluated in this study.

\section{Conclusion}
\label{conclusion}

This paper has examined the capabilities of LLMs in verifying concurrent programs under relaxed memory models. Our evaluation highlights both the potential and limitations of these models in addressing the inherent complexities of concurrency and memory consistency.

LLMs can be a powerful tool in the verification of concurrent programs, especially under related memory models. They can help developers understand and manage the complexities of concurrency, generate useful test cases, and even suggest optimizations. However, they should be used as a complement to, rather than a replacement for, traditional formal methods and testing tools, particularly when dealing with the nuanced and often complex nature of memory models.

This study underscores the potential and limitations of LLMs in analyzing concurrency issues within software programs. While GPT-4 outperforms other models in identifying data races and deadlocks under a sequentially consistent memory model, it, along with the other evaluated models, struggles to effectively verify program correctness under relaxed memory models like TSO and PSO. The findings highlight a significant gap in the current capabilities of LLMs, particularly in accurately capturing the complexities of memory ordering constraints. As concurrent programming continues to grow in importance, advancing LLMs to better handle these challenges is crucial for enhancing their reliability in concurrent program analysis and verification. Future research should focus on refining these models to address the nuanced demands of relaxed memory models, aiming to bridge the gap in their applicability to real-world concurrent programming scenarios.

\section{Availability}
\label{availability}

To promote reproducibility and enable further research, all resources and materials used in this study are fully accessible to the research community. This includes the datasets, such as SV-COMP's \textit{pthread} tests and ARM Litmus tests, as well as the complete code implementations used in our experiments. Furthermore, all custom scripts, evaluation metrics, and additional tools developed during this study are openly shared in a public repository, available at \url{https://anonymous.4open.science/r/LLM-Conc-RMM-13F8}.  This ensures that researchers can not only reproduce our results but also expand upon them by extending the scope of this work. By prioritizing transparency and accessibility, we aim to accelerate advancements in the analysis of concurrent programs under relaxed memory models and LLM evaluation.

\bibliographystyle{acm}
\bibliography{main}
\end{document}